\documentclass[12pt]{article}
\usepackage[cp1251]{inputenc}
\usepackage[T2A]{fontenc}

\evensidemargin\oddsidemargin \topmargin2cm

\usepackage{amsfonts,amscd,amsmath}

\usepackage{
amssymb,amsmath,
latexsym,euscript
}

\usepackage{amsfonts,amscd,amsmath}

\def\t {\tau }

\def\f {\varphi }
\def\e {\varepsilon }
\def\s {\sigma }

\def\a {\alpha }
\def\x {\xi }

\begin{document}

\title{\textbf{Averaging of one-parameter semigroups and\\ passage to the limit in the space of pseudomeasures}}

\author{V.Zh. Sakbaev $^1$\footnote{fumi2003@mail.ru} ,
 \\[2.7mm]
 ${}^1$\small{Moscow Institute of Physics and Technology
}\\
\small{Institutskiy per. 9, 141700, Dolgoprudny, Moscow reg., Russia}\\
}
\date{ }
\maketitle

\begin{abstract} 
The  sequence of one-parameter semigroups arising as the approximation of initial-boundary value problem with singularities is the object of investigation of this paper. The set of limit points of the sequence of approximating semigroups is studied. The set of limit points of the map with  values in a linear topological space is presented as the set of  mean values of this map by  measures on the domain of definition of the map.  
One to one correspondence betwin the semigroups generated by any  approximating  initial-boundary value problems and the pseudomeasures on the space of maps of time semiaxe into the coordinate space is studied. The linear space of pseudomesures endowed with the structure of Banach space and with the structure of the linear topological space such that the convergence of semigroup sequence is equivalent to the convergence of the sequence of  corresponding pseudomeasures. The desctiption of a limit point of the sequence of approximating semigroups is obtained by a measure on the topological vector space of corresponding pseudomeasures. The trajectories the limit one-parameter family of transformations of the space of initial data is described by the mean value of the random pseudomeasure.

\end{abstract}

\newpage

\section {Introduction}
The averagin of the set of elements of some linear space is the integral of identical map of the linear space onto itself by some nonnegative normalized measure with support on this set. The averaging procedure in the topological linear spaces of pseudomeasures or operator-valued functions is investigated. This procedure is applied to the description of approximation of Cauchy problem with the destruction (see \cite{Miz, Pan, BU, ZKO}) of solution by the sequence of Cauchy problems in some space of operators presenting the Cauchy problems. 

The notion of statistical solution was introduced in the papers by  \cite{Ars}, \cite{Vishi}, \cite{VL} for the investigation of the initial value problems wihtouth the uniqueness of solution. For studing of this class of problems the Cauchy problem for the maps of time interval into the space of measures on the space of initial data $H$ is considered instead of the Cauchy problem for the map of time interval into the space of initial data. The another transformation of the   Cauchy problem for the map of time interval into the space of initial data $H$ is to define the measure on the Banach space of solutions of Cauchy problem ($C(R_+,H)$ for example).

The initial-boundary value problem without the existence of solution can be transformed into the problem of description of the limit points set for the sequence of solution of approximating  initial-boundary value problems. One of the tools of the description of this limit points set is the measure on the space  of solution of approximating  initial-boundary value problems which presents the distribution of values of this sequence on the space  of solution, see
\cite{SR11}. 

The investigation of such limit mesures on the space of solutions had been studied in the paper  \cite{PV}. In this paper the barycenter of a limit mesure is described as the solution of some variational problem on the space of solutions. 
The procedure of continuation of the resolution map of the problem by the baricenter of limit measure on the space of solutions is similar to the  procedure of averaging (homogenization) of differential operators (see  \cite{ZKO}).

The procedure of averaging (homogenization) of differential operators is defined in \cite{ZKO} by the passage to the limit for the sequence of solutions of initial-boundary-value problems  depending on some smal parameters. In this book there is the posing of the new initial-boundary valued problem for some new differential equation such that the limit point of the sequence is the solution of this problem.  We show that the limit point of the sequence of solutions can be described as the barycenter of some measure on the space of solutions.

In this paper as in the work \cite{SSS} the map of the set of contactive semigroups of bounded linear operators of Hilbert space $H=L_2(R^d)$
into the space of pseudomeasures on the space of functions on time interval with values in the coordinate space $R^d$ is defined. In the space of pseudomeasures the structure of Banach space is introduced. 

The parametrization of the set of all limit points of the sequence of solutions of regularizing problems  by the family of nonnegative normalized measures on the set of regularizing problems is obtained.
The relation between the limit points of the sequence of regularizing semigroups and the limit points of the sequence of corresponding pseudomeasures is studied. It will be showed that the limit point of pseudomeasure sequence is the pseudomeasure but the limit point of the semigroup sequence can't be the semigroup. 

The parametrization of the of limit points  for the sequence of pseudomeasures is given by the set of normalised nonnegative measures on the elements of the sequence.
In particular this description includes the set of all generalised Banach limits of the the sequence  (see \cite{Semenov}, \cite{TF}) since any Banach limit of the sequence can be parametrized by the measure on the sequence elements which is invariant with respect to some group of its transformations.

The set of limit points of of the sequence of semigroups (of pseudomeasures) is endowed with the structure of the space with the measure by the introduction of the measure 
 on the elements of the sequence of approximating semigroups (pseudomeasures). This construction gives the opportunity for continuation of the solution of initialy-boundary value problem accross the moment of arising of singularities, bifurcation or blow-up by using of the random variable with values in the set of semigroups of transformation of the space of initial data (or by using ofthe random variable with values in the set of pseudomeasures on the set of maps of time intervale of the problem into its coordinate space). Thus the arising of the stochastic properties in the dynamical maps of initial-boundary value problem with the effect of destruction of solution or its nonuniqueness is based on the behavior of approximatioms of considered initial-boundary value problem in the space of such problems.

The structure of the paper is following. Firstly the theorem of presentation of the set of limit points of a map with values in topological vactor space by the set of mean values of this map with respect to arbitrary measures on the domain of this map has been proved. The second aim of this article is the relation between the semigroups of transformations of a Hilbert space and pseudomeasures on the space of maps of of time semiaxe into the coordinate space of the initial boundary value problem. The linear space of pseudomaesures is endowed with the topology and with the norm which corresponds to some toplogies in the space of operator-valued maps on time semiaxe. The third aim of the article is the random variables with the values in the topological vector spaces of pseudomeasures or operator-valued functions with different topologies. 
The relation between the measure on the space of pseudomeasures or on the space of operator-valued functions with the Young measures is considered (see \cite{Evans, NA}).

\section{ The limit points and the averaging of generalized sequences}.
Let $E$ be the topological space and $2^E$ is the algebra of all subsets of the set $E$. Let  $W(E)$ be the set of nonnegative normalized finite additive measures on the measurable space $(E,2^E)$. For any point $\e _0\in E$ the symbol $W(E,\e _0)$ notes the set of nonnegative normalized finite additive measures on the measurable space $(E,2^E)$ such that any measure  $\mu \in W(E,\e _0)$ is concentrated in arbitrary neighborhood of the point  $\e _0$ in the following sence: the equality $\mu (A)=0$ holds for any set $A\subset E$ such that $\e _0$ is not limit point of the set $A$. Let $W_0(E,\e _0)=\{ \mu \in W(E,\e _0):\ \mu (A)\in \{ 0,1\} \ \forall \ A\in 2^E\}$ be the subset of two-valued measures in the set $W(E,\e _0)$.

Let $S,Z$ be two linear spaces and $\beta :S\times Z\to C$ is the bilinear form $\beta :\ \beta (z,s)=\langle z,s\rangle =z(s)$ which defines the separated duality relation on the spaces $S,Z$  (see \cite{Shef}):
if $z(s)=0\ \forall \ s\in S$ then the equality $z=\theta _Z$ holds;
conversely, if $z(s)=0\ \forall \ z\in Z$, then $s=\theta _S$. 
Therefore weak topology $\s (Z,S)$ on the space $Z$ and weak topology $\s (S,Z)$ on the space $S$ are Hausdorff topology.

Linear functional $f$ on the space $S$  is continuous in $\s (S,Z)$ topology if and only if there is the unique element $z_f\in Z$ such that $f(s)=z_f(s)$ for any $s\in S$ 
 (see \cite{Shef}, p. 159, statement 1.2).
The reverce statement is true: the linear functional $f$ on the space  $Z$ is continuous in $\s (Z,S)$-topology iff 
there is the unique element  $s_f\in S$ such that $f(z)=\langle z,s_f \rangle $ for any $z\in Z$.

{\bf Theorem 1}. {\it Let $S,Z$ be some two linear spaces. Suppose that $\beta :S\times Z\to C$ is the bilinear form $\beta :\ \beta (z,s)=\langle z,s\rangle =z(s)$ which defines the separated duality relation on the spaces $S,Z$. Let $G:\ E \to Z$ be some map of topological space $E$ into the linear toplogical space $(Z,\s (Z,S))$. If the set $G(E)$ is bounded in the space 
$(Z,\s (Z,S))$ then
$$
{\rm Ls}_{\e \to \e _0}G(\e )=\bigcup\limits_{\mu \in W_0(E,\e _0)}\int\limits_EG(\e )d\mu ,\eqno (1)
$$
where ${\rm Ls}_{\e \to \e _0}G(\e )$ is the set of limit points of the map $G$ as $\e \to \e _0$ and integral in right hand side is Pittis integral
$\int\limits_EG(\e )d\mu =g\in Z\ \Leftrightarrow \ \f (g)=\int\limits_E\f (G(\e ))d\mu \ \forall \  \f \in S $.

If the limit $\lim\limits_{\e \to \e _0}G(\e )=g\in Z$ exists then $g=\int\limits_{E}G(\e )d\mu $ for any $\mu \in W(E)$.

In addition if the map $G$ is continuous in some deleted neighborhood of the point $\e _0$ and the set $E$ is arcwise connected then  
$$
{\rm Ls}_{\e \to \e _0}G(\e )=\bigcup\limits_{\mu \in W(E,\e _0)}\int\limits_EG(\e )d\mu .
$$

 }

Proof. For any functional $\f \in  S$ the function $\langle G(\e ),\f \rangle ,\, \e \in E,$ is bounded. Then for any measure  $\mu \in W(E)$ Pettis integral $I_{\mu ,G}(\f )=\int\limits_E\f (G(\e ))d\mu $ is defined and the map $I_{\mu ,G}$ is linear with respect to $\f $. Then the functional $I_{\mu ,G}$ is linear functional on the space $S$. It is obviously continuous in  $\s (S,Z)$ topology. Hence there is the unique element $g_{\mu ,G}\in Z$ such that $\f (g_{\mu ,G})=g_{\mu ,G}(\f )$ for all $\f \in {\cal S}$. 
Thus for any $\mu \in W(E)$ Pettis integral defines the unique point $\int\limits_EG(\e )d\mu =g_{\mu ,G}\in Z$.

If $\mu \in W(E,\e _0)$ then system of sets $\mu ^{-1}(1)\subset 2^E$ is the ultrafilter $\digamma _{\mu }$ of subsets of the set $E$. For any $F\in \digamma _{\mu }$ the point $\e _0$ is the limit point of the set $F$
in the topology  space  $E$.
Hence for any functional $\f \in S$ integral of the function %
$\f \circ G$ by measure $\mu $ is equal to the limit of the function $\f \circ G$ along the ultrafilter $\digamma _{\mu }$ since  for any $\delta >0$ the inclusion  $(\f \circ G)^{-1}(g_{\mu ,G}(\f )-\delta ,g_{\mu ,G}(\f )+\delta )\in \digamma _{\mu }$ holds according to the definition of Pettis integral.

The topology $\tau (Z,S)$ is generated by the family of neighborhoods  $\{ O_{\e ,\f }(g),\ \e >0,\, \f \in S,\, g\in Z\} $ which is the base of topology  $\tau (Z,S)$.

According to definition of the point  $g_{\mu ,G}$ for any $\e >0$ and any $\f \in S$ the inclusion  $G^{-1}[O_{\e ,\f }(g_{\mu ,G})]\in \digamma _{\mu }$ holds. Hence the limit of the map $G$ in the topology $\tau (Z,S)$ along the ultrafilter  $\digamma _{\mu }$ is equal to $g_{\mu ,G}$.
Therefore for any measure  $\mu \in W_0(E)$ the point $g_{\mu ,G}$ is the limit point of the map $G$  as  $\e \to \e_0$ i.e. $\bigcup\limits_{\mu \in W_0(E)}\int\limits_EGd\mu \subset {\rm Ls}_{\e \to \e_0}G(\e )$.

Let $g$ be the limit point of the map $G$ as %
$\e \to \e_0$. Hence for any $\f \in S$ and any  $\s >0$ there is the neighborhood $U_{\s ,\f }(\e _0)$ of the point $\e _0$ such that  $U_{\s ,\f }(\e _0)\subset G^{-1}(O_{\s ,\f }(g))$. The family of neighborhoods $\{ U_{\s ,\f }(\e _0),\ \s >0,\, \f \in S\}$ is the base of some filter $F$ of subsets of the set $E$.
 
Then if $\digamma $ is the ultrafilter majorizing the filter $F$ and  $\mu _{\digamma }$ is the corresponding two-valued measure then $g=\int\limits_{E}Gd\mu _{\digamma }$, therefore ${\rm Ls}_{\e \to \e_0}G(\e )\subset \bigcup\limits_{\mu \in W_0(E)}\int\limits_EGd\mu $. Therefore the equality (1) is proved.

If the limit  $\lim\limits_{\e \to \e _0}G(\e )=g\in Z$ exists then the set $
{\rm Ls}_{\e \to \e _0}G(\e )$ is one-point set and therefore $g=\int\limits_{E}G(\e )d\mu $  for any $\mu \in W(E)$. 

If in addition the set $E$ is arcwise connected and the map $G$ is continuous then the image $G(E)$ is arcwise connected. Therefore for any $\f \in S$ the set ${\rm Ls}_{\e \to \e _0}\f (G(\e ))$ is the segment. According to the equality $W_0(E)={\rm Extr}(W(E))$ any mesure $\mu \in W( E)$ is the limit point of convex hall of measures from the set $W_0(E)$. Hence for any $\f \in S$ the inclusion $\int\limits_E\f (G(\e ))d\mu \in {\rm conv}(\{ \int\limits_E\f (G(\e ))d\nu ,\, \nu \in W_0(E)\})={\rm Ls}_{\e \to \e _0}\f (G(\e ))$ holds.
Thus $\bigcup\limits_{\mu \in W(E)}\int\limits_E\f (G(\e ))d\mu \subset {\rm Ls}_{\e \to \e _0}\f (G(\e ))$ for any $\f \in S$ and hence $\bigcup\limits_{\mu \in W(E)}\int\limits_EG(\e)d\mu \subset {\rm Ls}_{\e \to \e _0}G(\e )$. Conversely according to the first statement of the theorem ${\rm Ls}_{\e \to \e _0}G(\e )=\bigcup\limits_{\mu \in W_0(E)}\int\limits_EG(\e)d\mu \subset \bigcup\limits_{\mu \in W(E)}\int\limits_EG(\e)d\mu$.

{\bf Remark 1.} If the topology $\s (Z,S)$ in the space $Z$ is generated by some total family of linear functionals ${\cal S}\subset S$ then the limit behavior of the map $G$ as $\e \to \e _0$ is defined by the family uf numerical functions $\{ f\circ G ,\, f\in {\cal S}\}$ on the set $E$.

\section{The semigroups in the Hilbert space $L_2(R^d)$ and corresponding pseudomeasures.
}
Let $D$ be some domain of finite dimentional ($d$-dimentional) euclidian space $R^d$ endowing with the Lebesque measure $\mu _L$. Let 
$L_2(D)$ be Hilbert space of square-integrable measurable by Lebesque complex-valued functions on the set $D$. In this paper we will asume that $D= R^d$.
The semigroups of transformations of the space $H=L_2(R^d)$ corresponding to the Cauchy problems for evolutionary equations (Shchrodinger equations or heat equations) are considered.

We obtaine that any one-parameter semigroup $\bf U$ of maps of the space $L_2(R^d)$ uniquely defines the pseudomeasure $\mu _{U}$ on the algebra $\cal A$ which is the minimal algebra conteining the cilindrical subsets of Banach space $C(R_+,R^d)$ (see \cite{SS2, SSS}). Here a pseudomeasure on the algebra $\cal A$ is the additive complex-valued function of set on algebra $\cal A$ without any assumption on its variation. The linear space of pseudomeasures on the algebra  $\cal A$ is noted by symbol ${\cal L}(C(R_+,R^d),{\cal A})$.

\subsection
{The classes of subsets}
Let ${\cal B}(R^{d})$ be the algebra of Borel set in $R^d$  and   ${\cal B}_{b}(R^{d})$ is the ring of bounded sets in algebra ${\cal B}(R^{d})$. Suppose that the family $ \Pi (R^d)$ of sets of algebra ${\cal B}(R^{d})$ is defined by the cundition:  for any element $A\in \Pi (R^d)$ either $A\in {\cal B} _b(R^d)$ or  $R^d\backslash A\in {\cal B} _b(R^d)$. Then the family $\Pi (R^d)$ is algebra.

The set $A^{t_{1},...,t_{m}}_{B_{1},...,B_{m}} ,\ t_1,...,t_m\in R_+,\, B_1,...,B_m\in {\cal B}(R^d)$ in the space $C(R_+,R^{d})$ is called cylindrical set iff it is the subset of the space $C(R_+,R^{d})$ which is defined by the finite number of conditions
$$
A^{t_{1},...,t_{m}}_{B_{1},...,B_{m}}=\{ \x \in C(R_{+},R^{d}):\, \x
(t_{j})\in B_{j} \, j\in \{1,...,m\} \}.\eqno (2)
$$
Here $m\in \bf N$ is arbitrary number and the series of $m$ nondecreasing numbers $t_{1},...,t_{m}\in R_+$ is arbitrary; and sets $B_{j}\in  {\cal B}(R^d)$ for any $j\in {\overline {1,m}}$. The finite series of the sets $B_j,\, j\in {\overline {1,m}}$ is called by the ($m$-dimentional) base  of cylindrical set (2).

The family of all cylindrical sets with arbitrary base $\{ B_{j}\in  {\cal B}(R^d),\, j\in {\overline {1,m}}\}$ (with bounded base $\{ B_{j}\in  {\cal B}_b(R^d),\, j\in {\overline {1,m}}$) is noted as symbol ${\rm Cyl}$ (is noted as symbol ${\rm Cyl}_b$). The symbol ${\rm Cyl}_2$ (the symbol ${\rm Cyl}_{2,b}$) notes the family of cylindrical sets with two-dimentional base $\{ B_{j}\in  {\cal B}(R^d),\, j\in {\overline {1,2}}\}$ (the family of cylindrical sets with two-dimentional bounded base $\{ B_{j}\in  {\cal B}_b(R^d),\, j\in {\overline {1,2}}\}$). Analogiously the symbol ${\rm Cyl}_{m}$ or ${\rm Cyl}_{m,b}$ notes the family of cylindrucal with $m$-dimentional base $\{ B_{j}\in  {\cal B}_b(R^d),\, j\in {\overline {1,m}}\}$.%

The family ${\cal P}$ of sets (2) such that for any $j\in {\overline {1,m}}$ the set $B_{j}$ belongs to the algebra
$\Pi (R^{d})$ is semialgebra of subsets of the space  $C(R_+,R^d)$. In fact, the family $\cal P$ conteines the unity $\Omega =C(R_+,R^d)$ and empty set. This family is closed with respect to intersections. For any set $A\in {\cal P}$ the complemetn $\bar A\equiv \Omega \backslash A$ is the union of finite number 
of sets from the family $\cal P$. For example if $B_1,B_2\in \Pi ,\  t_1,t_2\in R_+$ пїЅ $A^{t_2}_{B_2}, A^{t_1}_{B_1}\in {\cal P}$, then $A^{t_2}_{B_2}\bigcap A^{t_1}_{B_1}=A^{t_1,t_2}_{B_1,B_2}\in {\cal P}$; since $\bar B_1, \bar B_2\in \Pi $ then $\bar A^{t_1}_{B_1}=A^{t_1}_{\bar B_1}$ and $\bar A^{t_1,t_2}_{B_1,B_2}=A^{t_1,t_2}_{\bar B_1,\bar B_2}\bigcup A^{t_1,t_2}_{\bar B_1,B_2}\bigcup A^{t_1,t_2}_{B_1,\bar B_2}$. By induction we obtaine that for any set (2) the following relaion holds: ${\bar A}^{t_1,t_2,...,t_m}_{B_1,B_2,...,B_m}=A^{t_1}_{\bar B_1}A^{t_2,...,t_m}_{ B_2,..., B_m}\bigcup A^{t_1}_{B_1}\bar A^{t_2,...,t_m}_{B_2,...,B_m}\bigcup A^{t_1}_{\bar B_1}\bar A^{t_2,...,t_m}_{B_2,..., B_m}$.

Let ${\cal A}$ be the least algebra of sets of the space $C(R_+,R^d)$, conteining the semialgebra ${\cal
P}$ (any element of the algebra $\cal A$ is the union of finite number of elements of semialgebra  ${\cal P}$). 
Let ${\cal R}_{b}$ be the ring in algebra ${\cal A}$ such that for any $S\in {\cal R}_b$ the set $S$ is the union of finite number of sets $A^{t_1,...,t_m}_{B_1,...,B_m}$ of type (2) where   $B_j\in {\cal B} _b(R^d)$ for any $j\in  {\overline {1,m}}$. Let ${\cal A}_b$ be the least subalgebra of algebra $\cal A$ which conteining the ring ${\cal R}_b$. Then ${\cal A}= {\cal A}_b $ since any element $A\in {\cal A}$ can be obtained as the result of finite number of union, complemets and intersections of some finite series of sets of the ring ${\cal R}_b$.

\subsection
{On the generation of pseudomeasure by the semigroup} Any family of continuous linear operators
${\bf U}(t),\, t\in R_+,$ in the space $H$ uniquely defines the preudomeasure $\mu _{U}\in {\cal
L}(C(R_{+},R^d),{\cal A})$, which restriction onto the ring ${\cal R}_{b}$ is given by the equality 
$$
\mu _{U}(A^{t_{1},...,t_{m}}_{B_{1},...,B_{m}})=(\chi
_{B_{m}},{\bf U}(t_{m}-t_{m-1}){\bf P}_{B_{m-1}}{\bf
U}(t_{m-1}-t_{m-2}){\bf P}_{B_{m-2}}...{\bf U}(t_{2}-t_{1})\chi
_{B_{1}}).\eqno (3)
$$
Here ${\bf P}_{B}$ is orthogonal projector in the space $H$ acting as the multiplication on the indicator function of the set  $B$, and $\chi _{B}$ is the indicator function of the set $B$. Then the equality (3) defines the unique complex-valued additive function on the ring ${\cal R}_b$.

The function $\mu _U:\, {\cal R}_b\to {\bf C}$ can be uniquely continued from the  ring ${\cal R}_b$ onto the algebra  ${\cal A}$ by the normalised condition:
{\it if  $A\in {\cal A}$ such that $\Omega  \backslash A \in {\cal R}_b$ then
$$
\mu (A)=1-\mu (\Omega  \backslash A ).\eqno (4)
$$
}
The conditions (3), (4) are uniquely defines the additive comlex-valued function $\mu _U$ on the algebra ${\cal A}$ such that $\mu _U(\Omega )=1$. But the variation of pseudomeasure $\mu $ can be infinite.

For example,

1) if ${\bf U}(t)=e^{-it{\bf L}_{\Delta }},\, t\geq 0$ where ${\bf
L}_{\Delta}$ is Laplace operator in the space $R^{d}$ then relations (3), (4) define the pseudomeasure $\mu _{U_{-i\Delta }}$ which is coincides with the Feynman pseudomeasure on the ring ${\cal R}_b$ (see \cite{SSh}). But Feynman pseudomeasure has the infinite variation on the ring ${\cal R}_b$ and has no countable additive property. 

2) If ${\bf U}(t)=e^{t{\bf L}_{\Delta }},\, t\geq 0$
then the relations (3), (4) define the pseudomeasure  $\mu _{U_{\Delta }}$ which is coincides with the Winer measure on the algebra ${\cal A}$ (see \cite{SSh}). Winer measure is nonnegative normalized countable additive measure on the algebra ${\cal A}$.

\subsection
{On the pseudomeasures generating semigroups} Any pseudomeasure $\mu $ on the algebra ${\cal A}$, any series of increasing real nonnegative numbers $t_0,t_1,...,t_m$ and any series of sets $B_1,...,B_{m-1}\in {\cal B}(R^d)$ define the complex-valued function $\beta ^{t_0,t_1,...,t_m}_{\mu ;\ B_1,...,B_{m-1}}$ on the set $H\times H$. The value of the function $\beta ^{t_0,t_1,...,t_m}_{\mu ;\ B_1,...,B_{m-1}}$ on the ordered pair of indicator function $\chi _{B_0},\chi _{B_m}$ of the sets $B_0,B_m\in {\cal B}_b(R^d)$ is the complex number $\beta ^{t_0,t_1,...,t_m}_{\mu ;\ B_1,...,B_{m-1}}(\chi _{B_0},\chi _{B_m})=\mu (A^{t_0,t_1,...,t_m}_{B_0,B_1,...,B_m})$.
In particular if $m=1$ then the function $\beta _{\mu }^{t_0,t_1}$ has the value $\beta _{\mu }^{t_0,t_1}(\chi _{B_0},\chi _{B_1})=\mu (A^{t_0,t_1}_{B_0,B_1})$ on the ordered pair of indicator functions of the sets $B_0,B_1\in {\cal B}_b(R^d)$. 

Then the function $\beta ^{t_0,t_1,...,t_m}_{\mu ;\ B_1,...,B_{m-1}}$ can be extended onto sesquilinear form on the dense linear subspace  $H_S\equiv {\rm span}\{ \chi _B,\, B\in {\cal B}_{b}(R^d)\}$ of the space $L_2(R^d)$ by the rule of linearty:
for any ordered pair of linear combinations of indicator functions
$f=\sum\limits_{k=1}^mc_k\chi _{B_{0k}}$ and $g=\sum\limits_{j=1}^n\alpha _j\chi _{B_{1j}}$ the value  $\beta ^{t_0,t_1,...,t_m}_{\mu ;\ B_1,...,B_{m-1}}(f,g)$ is defined by the relation
$$
\beta ^{t_0,t_1,...,t_m}_{\mu ;\ B_1,...,B_{m-1}}(f,g)=\sum\limits_{k=1}^m\sum\limits_{j=1}^n c_k\bar \alpha _j\beta ^{t_0,t_1,...,t_m}_{\mu ;\ B_1,...,B_{m-1}}(\chi _{B_{0k}},\chi _{B_{1j}}).
$$


{\bf Definition 1a}. The pseudomeasure $\mu \in {\cal L}(C(R_+,R^d),{\cal A})$ is called continuous if for any increasing system of real nonnegative numbers $t_0,t_1,...,t_m$ and any  sets $B_1,...,B_{m-1}\in {\cal B}(R^d)$ the corresponded sesquilinear form $\beta ^{t_0,t_1,...,t_m}_{\mu ;\ B_1,...,B_{m-1}}$ on the space $H_S$ is bounded. I.e. there is the constant $M\geq 0$ such that the following inequality holds
$$
\| \beta ^{t_0,t_1,...,t_m}_{\mu ;\ B_1,...,B_{m-1}}\|=\sup\limits_{f,g\in H_S,\, \|f\|=1,\, \|g\|=1}|\beta ^{t_0,t_1,...,t_m}_{\mu ;\ B_1,...,B_{m-1}}(f,g)|
\leq M.
\eqno (5a)
$$

{\bf Definition 1as}. The pseudomeasure $\mu \in {\cal L}(C(R_+,R^d),{\cal A})$ is called continuous uniformly with respect to sets if for any increasing system of real nonnegative numbers $t_0,t_1,...,t_m$  the sesquilinear form $\beta ^{t_0,t_1,...,t_m}_{\mu ;\ B_1,...,B_{m-1}}$ on the space $H_S$ is bounded for any collection of the sets $B_1,...,B_{m-1}\in {\cal B}(R^d)$. I.e. there is the constant $M\geq 0$ such that the inequality (5a) holds for any collection of the sets $B_1,...,B_{m-1}\in {\cal B}(R^d)$.

{\bf Definition 1b}. The pseudomeasure $\mu \in {\cal L}(C(R_+,R^d),{\cal A})$ is called continuous with respect to Lebesgue measure if for any increasing system of real nonnegative numbers $t_0,t_1,...,t_m$ there is the constant $M\geq 0$ such that for any sets $B_0,B_1,...,B_m\in {\cal B}_b(R^d)$ the following inequality holds 
$$
\mu (A^{t_0,t_1,...,t_m}_{B_0,B_1,...,B_m})\leq M^m\Pi _{k=0}^{m}{\rm \mu }_{L} (B_k),\eqno (5b)
$$ 
where ${\rm \mu }_{L}$ is Lebesgue measure on $R^d$.

{\bf Definition 2a}. The pseudomeasure $\mu $ is called uniformly continuous  if there is the constant $M\geq 0$ such that for any increasing system of real nonnegative numbers $t_0,t_1,...,t_m$ and any sets $B_1,...,B_{m-1}\in {\cal B}(R^d)$ the inequality  (5a) holds.

{\bf Definition 2b}. The pseudomeasure $\mu $ is called uniformly continuous with respect to Lebesgue measure  if there is the constant $M\geq 0$ such that for any increasing system of real nonnegative numbers $t_0,t_1,...,t_m$ and any sets $B_0,B_1,...,B_m\in {\cal B}_b(R^d)$ the inequality  (5b) holds.

The condition of the definition 1as is the stronger than the condition of definition 1a, but any of it is no stronger and no weaker than the condition of the definition 1b. Let us consider the important example of Feynman pseudomeasure $\mu _F$ which is generated by the unitary group ${\bf U}(t)=e^{i\Delta t},\, t\in R,$ according to the equality (3). This pseudomeasure $\mu _F$ satisfies the conditions of the definition 2a with the constant $M=1$ in the inequality (5a). The conditions of the definition 1b are hold for any increasing system of real nonnegative numbers $t_0,t_1,...,t_m$ but the constant $M$ in the inequality 1b depends on this system of numbers $t_0,t_1,...,t_m$ and the conditions of the definition 2b are not hold.

If the pseudomeasure $\mu $ is continuous (in the sense of definition 1a) then the function 
$\beta _{\mu }^{t_0,t_1}$ 
can be uniquely continued onto the sesquilinear function on the space  $L_2(R^d)$. 
Analogiously for any increasing system of real nonnegative numbers $t_0,t_1,...,t_m$ and any sets $B_1,...,B_{m-1}\in {\cal B}(R^d)$ the function $\beta ^{t_0,t_1,...,t_m}_{\mu ;\ B_1,...,B_{m-1}}$ has the unique continuation onto the continuous sesquilinear functional on the space $L_2(R^d)$.

The continuous sesquilinear functional 
$\beta ^{t_0,t_1,...,t_m}_{\mu;\ B_1,...,B_{m-1}}$ 
uniquely defines the continuous linear operator ${\bf A} ^{t_0,t_1,...,t_m}_{\mu ;\ B_1,...,B_{m-1}} \in B(L_2(R^d))$ such that 
$$
\beta  ^{t_0,t_1,...,t_m}_{\mu ;\ B_1,...,B_{m-1}} (u,v)=({\bf A} ^{t_0,t_1,...,t_m}_{\mu ;\ B_1,...,B_{m-1}}u,v)_{L_2}\, \forall \, u,v\in L_2(R^d).\eqno (6)
$$ 
In particular if $m=1$ then the continuous pseudomeasure $\mu $ defines the two-parametric family of operators  ${\bf A} ^{t_0,t_1}_{\mu },\, t_0,t_1\in R_+.$

{\bf Definition 3}. The continuous pseudomeasure $\mu $ is called Markovian if the family of operators   
${\bf A}^{t_0,t_1,...,t_m}_{\mu ;\ B_1,...,B_{m-1}},\ t_0,t_1,...,t_m\in R_+,\ B_1,...,B_{m-1}\in {\cal B}(R^d),$  satisfies the conditions:

1) Markov condition (evolutionary condition): 
$${\bf A}^{t_0,t_1,...,t_m}_{\mu ;\ B_1,...,B_{m-1}}{\bf P}_{B_m}{\bf A}^{t_m,t_{m+1},...,t_{m+p}}_{\mu ;\ B_{m+1},...,B_{m+p-1}}={\bf A}^{t_0,...,t_{m+p}}_{\mu ;\ B_1,...,B_{m+p-1}}$$ $$\quad \forall \, t_0\leq t_1\leq ...\leq t_{m+p}\in R_+\quad \forall \, B_1,...,B_{m+p-1}\in {\cal B}(R^d);\eqno (7)
$$

2) the normalise condition:
${\bf A} _{\mu }^{t_0,t_0}={\bf I}$ for all $t_0\in R_+$.

If the pseudomeasure $\mu $ is continuous and Markovian then it defines by means of equality (6) the two-parametric family of operators ${\bf A}_{\mu }^{t_1,t_2},\, t_1,t_2\in R$, which  satisfies the equality: 
$$
{\bf A}_{\mu }^{t_1,t_2}{\bf A}_{\mu }^{t_2,t_3}={\bf A}_{\mu }^{t_1,t_3}\, \forall \, t_1,t_2,t_3\in R_+:\ t_1\leq t_2\leq t_3.\eqno (8)
$$

{\bf Remark 2.} Markovian property of continuous pseudomeasure $\mu $ gives the opportunity to    difine the pseudomeasure $\mu $ on the algebra ${\cal A}$ by its restriction onto the class of subsets  $Cyl_{2}$.

{\bf Definition 4}. The pseudomeasure $\mu $ is called stationary if it is invariant with respect to the shift of all time parameters on some real variable:
$$
\mu (A_{B_0,...,B_m}^{t_0,...,t_m})=\mu (A_{B_0,...,B_m}^{t_0+s,...,t_m+s}) \eqno (9)
$$
for any $s\in R_+,\, A^{t_0,...,t_m}_{B_0,...,B_m}\in {\rm Cyl}$.

{\bf Remark 3.} { If the Markovian pseudomeasure $\mu
$ on the algebra ${\cal A}$ is stationary then the equalities (6) defines the two-parametric family of operators ${\bf A}_{\mu }^{t_1,t_2}$ such that it dependes only on the difference $t_2-t_1:$ 
$$
{\bf A}_{\mu }^{s,t+s}={\bf V}_{\mu }(t),\, t\in R_+,\quad \forall \ s\in R_+.\eqno (10)
$$
Then the pseudomeasure $\mu $ defines the one-parametric semigroup  ${\bf V}_{\mu }(t),\, t\in R_+$ in the space $H$.
The presentation and the approximation of the semigroup generating by the Cauchy problem for the evolution equation can be given by the means of Feynman formulas (see \cite{BSS}).
}

 {\bf Remark 4}. If the pseudumeasure $\mu $ is Markovian but is not stationary then the equality (6) defines the two-parametric family of bounded linear operators
 ${\bf A}_{\mu }^{s,t},\, 0\leq s\leq t<+\infty ,$ which is called evolutionary family. This evolutionary family describes the dynamics of a solution of Cauchy problem with time depended Hamiltonian.

 {\bf Remark 5.} If continuous pseudomeasure is not markovian then the family of operators ${\bf U}_{\mu }^{s,t},\, 0\leq s\leq t<+\infty ,$
has no evolutionary property (8). According to the works
 \cite{STMF}, \cite{IDA} any limit point for the sequence of regularised semigroups in the space of quantum states is the family of dynamical maps which  has no evolutionary property. 

 {\bf Remark 6}. If the pseudomeasure $\mu $ defines the diffusion random process with value in the space $R^{d}$, then the generator the semigroup ${\bf U}_{\mu }$ obtained by the equation  (3) is linear second order elliptic diferential operator.

{\bf Proposition 1}. {\it If the pseudomeasure $\mu _{U}$  is generated by the group of unitary operators ${\bf U}(t),\, t\in  R,$ according to equality (3) then the pseudomeasure $\mu _{U}$ possesses Markovian, continuity and stationarity properties}.
 
In fact, for all $T>0$ the following estimate holds
 $$
 \sup\limits_{|t|<T,\, v,u\in S_{2},\, \|v\|_{H}=1,\,
 \|u\|_{H}=1}|(u,{\bf V}_{\mu }(t)v)|\leq 1.
 $$
Therefore pseudomeasure $\mu _U$ satisfy the continuity condition (5) with the constant $M=1$.
According to the equality (3) the Markovian property and stationarity property of pseudomeasure  $\mu _U$ is the concequence of semigroup properties of the family of operators ${\bf U}(t),\, t\in R$.

 {\bf Corollary 1}. If the pseudomeasure $\mu _{U}$ is generated by the semigroup ${\bf U}(t),\, t\in R$ according to the equality (3) then the semigroup can be uniquely reconstructed by pseudomeasure $\mu _U$ according to the equality (10). Conversely if the semigroup ${\bf V}_{\mu }$ is defined by the continuous Markovian stationary pseudomeasure $\mu $ according to the equalities (10), then the semigroup ${\bf V}_{\mu }$ uniquely defines the pseudomeasure $\mu $  by the equalities (3).

 Therefore we can identify 
the one-parametes semigroups of operators and the stationary Marcovian continuous pseudomeasures.
Hence the passage to the limit for the sequences of solutions of the Cauchy problems in the space of solutions $C(R_+,L_2(R^d))$ is corresponded to the passage to the limit for the sequences of pseudomeasures on the space $C(R_+,R^d)$ in the space of pseudomeasures ${\cal L}(C(R_+,R^d),{\cal A})$.  

Let symbol ${\cal L}_{cont}(C(R_+,R^d),{\cal A})$ (or symbol ${\cal L}_{ucont}(C(R_+,R^d),{\cal A})$) note the set of pseudomeasures in the space ${\cal L}(C(R_+,R^d),{\cal A})$ which is continuous (or uniformly continuous). Both of this sets are the linear subspaces in the space  ${\cal L}(C(R_+,R^d),{\cal A})$.

\subsection{On the topologies on the pseudomesures space.} 
The linear space ${\cal L}(C(R_+,R^d),{\cal A})$ can be endowed with the different topologies. 
The article \cite{SF} is devoted to the description of wide class of this topologies, to the investigation of compantness and convergence for the sequences of Radon measures on the completely regular topological spaces, to investigation of the properties of limit measures.

Let us introduce the equivalence relation on the space of pseudomeasures ${\cal L}(C(R_+,R^d),{\cal A})$. The pseudomeasure $\mu \in {\cal L}(C(R_+,R^d),{\cal A})$ is called equivalent to the pseudomeasure $\nu \in {\cal L}(C(R_+,R^d),{\cal A})$ if the equality $\mu (A)=\nu (A)$ holds for any set $A\in {\rm Cyl}_{b}$. For examle if the pseudomeasure $\mu $ is defined by the unitary group $\bf U$ according to the equalities (3), (4), and pseudomeasure $\nu $ is defined by the same group $\bf U$ according to the equality (3) and the equality $\mu _{{\bf U}}(A)=-\mu _{\bf U}(\Omega \backslash A)$, $\Omega \backslash A\in {\cal R}_b$  instead of the condition (4) then any of this two pseudomeasures are equivalent to each other.

Let symbol 
${\cal L}_{ucL}$ note the set of uniformly continuous with respect to Lebesgue measure pseudomeasures on the algebra ${\cal A}$ (see definitions 1b, 2b). Then the set  ${\cal L}_{ucL}$ is the linear subspaces in the space ${\cal L}(C(R_+,R^d),{\cal A}
)$.
 
{\bf Lemma 1}. {\it  Suppose that the map $p:\  {\cal L}_{uc}\to [0,+\infty )$ has value $p(\nu )=m_{\nu}$ on arbitrary pseudomeasure $\nu \in {\cal L}_{uc}$ which is equal to the least constant $M_{\nu }$ in the inequality (5b) (see definition 2b). Then the functional $p$ is the norm on the linear space ${\cal L}_{ucL}$.}

In fact, the functional $p$ is defined on the space ${\cal L}_{u}$ and has value $p(\nu )\in [0,+\infty )$ at any point $\nu \in {\cal L}_{uc}$. 
The functional $p$ on the space ${\cal L}_{uc}$ is obviously nonnegative and uniform. The triangle inequality $m_{\nu _1+\nu _2}\leq  m_{\nu _1}+m_{\nu _2}$ is the concequence of the definition 1b and the definition of summ of pseudomeasures. If $p(\mu )=m_{\mu }=0$ for some pseudomeasure $\mu \in {\cal L}_{uc}$ then according to the definition 1b $\mu (A)=0$ for any $A\in {\rm Cyl}_{b}$, hence $\mu =\theta _{{\cal L}_{ucL}}$.

{\bf Theorem 2}. {\it Normalised linear space $({\cal L}_{ucL},p)$ is Banach space.}

Let $\{ \nu _k\}$ be fundamental sequence of pseudomesures in the normalised space $({\cal L}_{ucL},p)$:  $\sup\limits_{n\in {\bf N}}p(\nu _k-\nu _{k+n})\to 0$ as $k\to \infty$. Therefore for any $A^{t_0,...,t_m}_{B_0,...,B_m}\in {\rm Cyl}_{b,m+1}\subset {\rm Cyl}_{b}\subset {\cal A}$ the inequality  $|\nu _k(A)-\nu _{k+n}(A)|\leq p(\nu _k-\nu _{k+n})^m\Pi _{j=0}^m\mu _L(B_j)$ holds. Then according to definition 1b the numerical sequence $\{ \nu _k(A)\}$ is fundamental for any $A\in {\cal A}$.  
Hence the equality
$$
\nu (A)=\lim\limits_{n\to \infty }\nu _n(A) \ \forall \ A\in {\cal A}\eqno (11)
$$ 
uniquely defines the function  $\nu (A),\, A \in {\cal A}$
on the algebra  ${\cal A}$.

Since the function $\nu _n$ on the algebra ${\cal A}$ is additive for any $n\in \bf N$, then according to (11) the function $\nu $ on algebra ${\cal A}$ is additive and hence 
$\nu \in {\cal L}(C(R_+,R^d),{\cal A})$. Let us prove that $\nu \in {\cal L}_{ucL}(C(R_+,R^d),{\cal A})$.  For any $n\in \bf N$ there is the constant $M_n=p(\nu _n)$ such that the inequality $|\nu (A)|\leq M_n^m\Pi _{j=0}^m\mu _L(B_j)$ holds for any $A^{t_0,...,t_m}_{B_0,...,B_m}\in {\rm Cyl}_{b,m+1}\subset {\rm Cyl}_{b}\subset {\cal A}$. The sequence $\{ M_n\}$ is bounded according to the fundamentality of the sequence $\{ \nu _n\}$ in the space $({\cal L}_{ucL},p)$. Therefore there is the constant $M\geq 0$ such that the inequality  $|\nu (A)|\leq M^m\Pi _{j=0}^m\mu _L(B_j)$ holds for any  $A^{t_0,...,t_m}_{B_0,...,B_m}\in {\rm Cyl}_{b,m+1}\subset {\rm Cyl}_{b}\subset {\cal A}$. Theorem 2 is proved.

\bigskip

Further some topologies on the linear space  ${\cal L}(C(R_{+},R^{d}),{\cal A})$ will be introduced.
 Let symbol $\s _{cyl}$ note the topology on the linear space ${\cal L}(C(R_{+},R^{d}),{\cal A})$ which is defined by the family of cylindrical linear functionals 
$$\{ P_{A},\, A\in {\cal A}\},\eqno (12)$$
where $P_{A}(\mu )=\mu (A)$.

Let $\sigma _{cyl,2}$ be the topology on the linear space  ${\cal L}(C(R_{+},R^{d}),{\cal A})$ which is defined by the linear functionals  $P_A,\, A\in Cyl_2$.
It is easy to show that the topology $\sigma _{cyl,2}$ is more weak than the topology $\sigma _{cyl}$.
 
Let $H_{S}$ be the pre-Hilbert space of step complex-valued functions on the space $R^d$ (i.e. the linear hall of the set of indicator functions for borel subsets $B\in {\cal B}(R^d)$) which is endowed with the scalar product of the space $H$ (see definition 1a).
 
 For any pair of functions $v,w\in H_S$ and any pair of numbers $s,t>0,\,
 s<t$, the functional $P_{s,t,v,w}$ on the pseudomeasures space ${\cal L}(C(R_{+},R^{d}),{\cal A})$ is defined by the rule: 

if the functions $v,w$ are indicator functions of Borel subsets $V,W\in {\cal
 B}(R^d)$ respectively then the value $P_{s,t,v,w}(\mu ),\, \mu \in {\cal L}$, is equal to the value $\mu (A^{s,t}_{V,W})$ of pseudomeasure  $\mu $ on the cylindrical set with two-dimentional base $A^{s,t}_{V,W}=\{ \xi:\ \xi
 (s)\in V,\, \xi (t)\in W \}\subset {\cal A}$. 

for linear combination of indicator function the the value of the functional $P_{s,t,v,w}(\mu )$ on the pseudomeasure $\mu $ is obtained by linearity.

Then for any Markovian continuous pseudomeasure
$\mu $ the following equality holds
 $$
 P_{s,t,v,w}(\mu )=|\int\limits_{C(R_{+},R)}(v(\xi (s)),w(\xi (t)
 ))d\mu |=|(w,{\bf U}(t-s)v)|=\mu (A^{s,t}_{V,W}).\eqno (13)
 $$

 Hence the topology on the linear space ${\cal L}(C(R_+,R^d),{\cal A})$ which is defined by the set of functionals (13) coincides with the topology $\s _{cyl,2}$ which is generated by cylindrical sets wih two-dimentional bases. Therefore it more weak than the topology which is generated by all cylindrical subsets.
 
To the investigation of uniform convergence of the sequence of pseudomeasures we introduce the topology  $\Sigma _{cyl,2}$ on the linear space  ${\cal L}(C(R_+,R^{d}),{\cal
 A}_b)$ which is generated by the system of nonlinear functionals $\{ P_{v,T},\
 T>0,\, v\in H\}$  where each seminorm $P_{v,T}$ is given by the formula:
 $$
 P_{v,T}(\mu )=\sup\limits_{t\in [0,T],\, w\in H_{S},\, \|w\|=1}P_{0,t,v,w}(\mu);
 \quad \ \mu \in {\cal L}(C(R_+,R^{d}),{\cal A}).\eqno (14)
 $$

The topology on the linear space ${\cal L
}(C(R_{+},R^{d}),{\cal A})$ which is defined by the system of functionals
$P_{s,t,v,w},\, s,t>0,\, v,w\in H,$ is corresponded to the topology on the space $C_s(R_+,B(H))$ of strongly continuous operator-valued functions which is generated by the system of functionals $p_{s,t,v,w},\, s,t>0,\, v,w\in H,$ in the following sence: for any pseudomeasure  $\mu \in {\cal L}(C(R_{+},R^{d}),{\cal A})$ the operator-function ${\bf U}_{\mu }(s,t),\, s,t>0$ arising acording to the equality (10) satisfy the equalities
$$
P_{s,t,v,w}(\mu )=|\int\limits_{C(R_{+},R^{d})}(v(\xi (s)),w(\xi
(t) ))d\mu |=|(w,{\bf U}_{\mu }(s,t)v)|=p_{s,t,v,w}({\bf U}_{\mu
}(t))
$$
for any pair of numbers $s,t>0$ and any pair of elements $v,w\in H$.

Analogiously for any strongly continuous semigroup ${\bf U}\in C_{s}(R_{+},B(H))$ the pseudomeasure $\mu \in {\cal L}(C(R_{+},R^{d}),{\cal A})$ arising according to the equality (3) 
(or if the stationary Marcovian continuous pseudomeasure $\mu $ defines the semigroup $\bf U$ according to equality (10)), 
satisfy the equalities
$$
P_{T,v}(\mu )=p_{T,v}({\bf U}),\quad T>0,\, v\in
H.\eqno (15)
$$

The following statement is the concequence of the equalities (15). The formulation of this statement and the scheme of the proof are published in the paper \cite{SS2}. Now we give the proof with detales  of this statement.

{\bf Theorem 3.} {\it Let $\{ \mu _{n}\}$ be the sequence of Markovian pseudomeasures on the algebra ${\cal A}$ such that for any  $n\in \bf N$
the pseudomeasure $\mu _{n}$ is uniformly continuous.
Suppose that the pseudomeasure $\mu _{n}$ defines the unitary semigroup ${\bf U}_{n}(t),\,
t\in R_+,$ in the space $H=L_{2}(R^{d})$ according to the equality (10). Any  semigroup  ${\bf U}_{n}(t),\,
t\in R_+,$ has the self-adjoint generators  ${\bf L}_{n}$. 

Then the convergence of the sequence of semigroups in the strong operator topolgy of the space $B(H)$ uniformly on arbitrary segment of semiaxe  $R_+$ to the limit operator-valued function ${\bf F}$ is equivalent to the convergence of the sequence of pseudomeasures  $\mu _{n}$ to the limit function of set $\nu $ on the algebra  ${\cal A}$ in the topology $\Sigma _{cyl,2}$.

If one of two equivalent conditions is fullfield then

1. The limit operator-function ${\bf F}$ is semigroup.

2. The sequence  $\mu _n$ converges to the limit function of a set $\nu \in {\cal L}$ in the topology $\s _{cyl}$. The limit function of a set $\nu $ is Markovian, uniformly continuous and stationary pseudomeasure.

3. The limit pseudomeasure  $\nu $ defines the semigroup  ${\bf F}(t)$ according to the equality 
(10). 
The limit pseudomeasure  $\nu $ can be defined by the semigroup ${\bf F}(t)$ according to the equalities
$$
\nu (A)=\lim\limits_{n\to \infty }\mu _{n}(A)=(\chi _{B_{m}},{\bf
F }(t_{m}-t_{m-1}){\bf P}_{B_{m-1}}{\bf
F}(t_{m-1}-t_{m-2})...{\bf P }_{B_{2}}{\bf F}(t_{2}-t_{1})\chi
_{B_{1}}),\eqno (16)
$$
which satisfy for arbitrary set
$A=A^{t_{1},...,t_{m}}_{B_{1},...,B_{m}}\in {\cal A}_{f}$.}

{\bf Proof.} According to (15) for any $n,k\in \bf N$
$T>0,\, v\in H$ the equality $P_{T,v}(\mu _{n}-\mu
_{k})=p_{T,v}(U_{n}-U_{k})$ holds. Therefore if the sequence of pseudomeasures $\{ \mu _{n} \}$ converges in the topology generated by the functionals (14)
to the limit function of a set $\nu $ on the class of subsets ${\rm Cyl}_2$, then the semigroup sequence $\{ {\bf U}_{n}\}$ converges in the strong operator topology uniformly on arbitrary segment to the limit operator-function ${\bf U}(t)$. The limit operator function is semigroup (see \cite{SR11}, \cite{SS2}). Hence the equality $(\chi _{B_0},{\bf F}(t)\chi _{B_1})=\nu (A^{0,t}_{B_0,B_1})$ is the consequence of  the  sequence of equalities  $(\chi _{B_0},{\bf U}_n(t)\chi _{B_1})=\mu _n(A^{0,t}_{B_0,B_1}),\, n\in \bf N$. Therefore the semigroup $\bf F$ is generated by the limit pseudomeasure  $\nu $ in accordance with the equality (10).

Conversely if the sequence of unitary semigroups $\{ {\bf U}_{n}\}$ converges in the strong operator topology uniformly on arbitrary segment to the limit operator function ${\bf U}(t),\, t\in R _+,$ then the limit function $\bf U$ is isometric semigroup and the correspondind sequence of pseudomeasures $\{ \mu _{n}\}$ converges in the topology generated by the functionals (14) 
to the limit function of a set $\nu (A),\, A\in Cyl_{2}$. 

Let us prove that if the sequence of continuous Markovian pseudomesures converges in the topology $\Sigma _{cyl2}$ to the function of a set $\nu $ on the class of subsets $Cyl_2$ then it converges in the topology generated by the functionals of a class $Cyl _{b}$ to the limit Markovian pseudomeasure $\cal V $ on the algebra ${\cal A}$ such that its restrictions on the class $Cyl_{2,b}$  coincides with the function $\nu $ and moreover the limit pseudomesure  $\cal V $ and limit semigroup  $\bf F$ satisfy the equalities  (10) and (16).

According to the theorem assumption any pseudomeasure  $\mu _{n},\, n\in \bf N$, is Markovian (see definition 3 and equality (7)), therefore for any $t,s>0$ and any $V,W,Y\in {\cal B}_{f}$ the following equalities
$$
\mu _{n}(A^{0,t,t+s}_{V,W,Y})=(\chi _{Y},{\bf U}_{n}(s){\bf
P}_{W}{\bf U}_{n}(t)\chi _{V})=({\bf U}_{n}(-s)\chi _{Y},{\bf
P}_{W}{\bf U}_{n}(t)\chi _{V})\eqno (17)
$$
holds. To obtaining the last equation we use the equality  $({\bf  U}_{n}(s))^*={\bf U}_{n}(-s),\, s\in
R_{+},\, n\in {\bf N},$ (where ${\bf
U}_{n}(s)=e^{-is{\bf L}_{n}}$) which is the consequence of self-adjointness of  operators 
${\bf L}_{n},\, n\in {\bf N}$ with arbitrary $s\geq 0$.

According to the first statement of the theorem the convergence of  semigroup sequence  ${\bf
U}_{n}(t)$ in the strong operator topology uniformly on any segment is equivalent  to convergence of the sequence$\{ \mu _{n}\}$ of continuous Markovian pseudomesures in the topology $\Sigma _{cyl2}$ to the limit function of a set on the class ${\rm Cyl}_2$. 
Then according to the convergence of semigroup sequence  ${\bf
U}_{n}(t)$ in the strong operator topology uniformly on any segment the following statement holds. The sequences of $H$-valued functions $\{ {\bf U}_{n}(t)\chi _{V}\}$
and $\{ {\bf U}_{n}(-s)\chi _{Y}\}$ converges uniformly on arbitrary segment in the space $L_{2}(R^{d})=H$ to the functions ${\bf U}(t)\chi _{V}$ and ${\bf
U}(-s)\chi _{Y}$ respectively. 
Therefore the limit  function of a set $\nu $ satisfies the equalities $\nu
(A^{0,t,t+s}_{V,W,Y})=(\chi _{Y},{\bf U}(s){\bf P}_{W}{\bf U}(t)\chi
_{V})$. Hence it satisfies the equality (7) for arbitrary pair of cylindrical subsets with two-dimentional base.

For any $n\in \bf N$ the pseudomeasure $\mu _{n}$ is generated by the semigroup ${\bf U}_n(t),\, t\geq 0,$ according to the equality (3). Therefore fore arbitrary increasing system of nonnegative numbers $t_0,t_1,...,t_m$ 
and any system of Borel sets $B_1,...,B_{m-1}\in {\cal B}_b(R^d)$ the pseudomeasure $\mu _{n}$ defines the sesquilinear forms  $\beta_{\mu _n;\ B_1,...,B_{m-1}}^{t_0,t_1,...,t_m} $
and bounded operators (6).
Since the pseudomeasure $\mu _n$ satisfyes the equality (3) then the operator (6) admites the following presentation ${\bf A}_{\mu _n;\ B_1,...,B_{m-1}}^{t_0,t_1,...,t_m}={\bf U}_{n}(t_1-t_0){\bf P}_{B_1}...{\bf P}_{B_{m-1}}{\bf U}_n(t_m-t_{m-1})$. Then according to the convergence of semigroup sequence $\{ {\bf U}_n\}$ in the strong operator topology uniformly on arbitrary segment     the sequence of operators   $\{ {\bf A}_{\mu _n;\ B_1,...,B_{m-1}}^{t_0,t_1,...,t_m}\}$ converges in the strong operator topology to the limit operator $ {\bf A}_{ B_1,...,B_{m-1}}^{t_0,t_1,...,t_m}={\bf F}(t_1-t_0){\bf P}_{B_1}...{\bf P}_{B_{m-1}}{\bf F}(t_m-t_{m-1})$ for arbitrary increasing system of nonnegative numbers $t_0,t_1,...,t_m$ 
and any system of Borel sets $B_1,...,B_{m-1}\in {\cal B}_b(R^d)$.     
Hence the sequence of functions $\{ \mu _n\}$ on the ring  ${\cal R}_b$ converges point-vice on the ring ${\cal R}_b$ to the function of set $\cal V$ such that the equality   $$
{\cal V}({ A}_{ B_0,B_1,...,B_{m}}^{t_0,t_1,...,t_m})=(\chi _{B_0},{\bf F}(t_1-t_0){\bf P}_{B_1}...{\bf P}_{B_{m-1}}{\bf F}(t_m-t_{m-1}) \chi _{B_m})\eqno (18)
$$ 
holds for any increasing system of nonnegative numbers $t_0,t_1,...,t_m$ 
and any system of Borel sets $B_1,...,B_{m-1}\in {\cal B}_b(R^d)$.  Hence
 the function of a set $\cal V$ is  the continuation of the function $\nu $ from the class $Cyl_2$ onto the ring ${\cal R}_b$ and has the unique continuation onto the measure $\mu _{F}$ on the algebra ${\cal A}$. Thus according to (18) the limit pseudomeasure $\mu _{{\bf F}}$ is defined by the limit semigroup $\bf F$ in accordance with the equality (3), (4) and the equality (16) is fulfield, i.e. ${\cal V}=\mu _{{\bf F}}$.

If $\mu _{F}$ is stationary Markovian pseudomeasure on the algebra ${\cal A}$ which is defined by the semigroup ${\bf F}$ in accordance with the rule 
(17) then the pseudomeasure $\mu _{F}$ satisfies the uniformly continuosn
condition (5a) with the constant $M=1$ due to the unitarity of semigroup  ${\bf F}$. Moreover for any  $T>0,\, v\in H$ the equality $\lim\limits_{n\to \infty }P_{T,v}(\mu _{n}-\mu
_{F})=0$ holds.

Since any of semigroups generating by the pseudomeasures $\mu _{n}$ is unitary then any of the constants $M^{n}$ in the conditions (5a) of pseudomeasure $\mu _{n}$ continuation can be equal to unity. 
ince the limit pseudomesure $\nu $ is defined by the limit semigroup $\bf F$ in accordance the equalities (18) then pseudomeasure $\nu $ is also uniformly continuous with the constant $M=1$.
Hence the limit pseudomeasure $\nu $ generates the semigroup ${\bf
F}$ according to the equality  (3) and can be obtained by using of semigroup ${\bf F }$ in accordance with the equality (16). Theorem 3 is proved.

 {\bf Corrolary 2}. The set  ${\cal L}_{sg}(C(R_+,R),{\cal A})$ of pseudomeasures generating a semigroups is closed in the topology $\Sigma _{cyl 2}$.


But the convergence of semigroups sequence in more weak topology can not give the information about the convergence of corresponding pseudomeasures sequence  on the hole algebra ${\cal A}$.
Let us note by symbol  ${\cal M}_a (C(R_{+},R^{d}),Cyl _{2})$ the linear space of additive functions on the class of subsets 
$Cyl _{2}$ (which is not closed with respect to finte number of intersection and unity operations). Let ${\cal W }=\{ P_{0,t,v,w}, t>0,\, v,w\in H\}$ be the set of linear functionals on the linear space  ${\cal M}_a (C(R_{+},R^{d}),Cyl _{2})$ acting by the formular (13) (see \cite{SSS}).
Let $\tau _{W}$ be the topology on the space  ${\cal M}_a (C(R_{+},R^{d}),Cyl _{2})$ which generated by the set of functionals $\cal W$. Then the convergence of  pseudomeasures sequence in the topology $\tau _W$ is equivalent  to the convergens of corresponding operator-valued functions sequence in the weak operator topology (see \cite{SSS}).

 {\bf Theorem 4.} {\it 
Let $\{ {\bf U}_{n}(t),\, t\geq 0\}$ be the sequence of unitary semigroups acting in the space $H=L_{2}(R^{d})$, and $\{ \mu ^{{\bf U}_n}\}$ is the sequence of additive functions on the class ${\rm Cyl}_{2 ,b}$ genetated by the semigroups  ${\bf U}_{n}(t),\, t\geq 0$ according to the equality (3).

Then the point-vice convergence on the semiaxe $R_{+}$
of the sequence of semigroups $\{ {\bf U}_{n}(t),\, t>0\}$
in the weak operator topology to the limit operator-valued function ${\bf F}(t)$  is equivalent to the convergence of functions of the set  sequence $\{ \mu _{n}\} $ in the topology $\tau _W$ to the function of the set  $\nu \in {\cal M}_a
 (C(R_{+},R^{d}),Cyl _{2})$
which is defined by the operator-valued function $\bf F$ by the equalities 
$$\nu
 (A)=\lim\limits_{n\to \infty }\mu _{n}(A)=(\chi _{B_{2}},{\bf F
 }(t_{2}-t_{1})\chi _{B_{1}}),\ \forall \ A=A^{t_{1},t_{2}}_{B_{1},B_{2}}\in
 Cyl_{2} .\eqno (19)
$$}
 
In fact if the sequence of the semigroups $\{ {\bf
 U}_{n}\}$ converges to the limit operator valued function ${\bf F}(t),\, t>0,$ in the weak operator topology pointwice on the semiaxe $R_{+}$, then the sequence of pseudomeasures $\{ \mu ^{{\bf U}_n}\} $ converges in the topology $\tau _W$ to the limit function of a set $\nu (A),\,
 A\in Cyl_{2} $ which is defined by the limit operator valued function ${\bf
 F}(t),\, t>0,$ according to the equality (19).
 
Conversely, suppose that the sequence of the pseudomeasures $\{ \mu _{n}\}$ is defined by the sequence of strongly continuous semigroups ${\bf U}_{n}$ by the equalities (3). If the sequence $\{ \mu _n\}$ converges in the topology $\tau _{{\cal W}}$
to the limit fonction of a set $\nu \in {\cal M}_a
 (C(R_{+},R),Cyl _{2})$ on the class of the sets  $Cyl_{2}$ then the sequence of the semigroups $\{ {\bf U}_n\}$ converges in the weak operator topology point-vese on the semiaxe $R_+$ to the limit operator valued function ${\bf F}(t)$ which is weakly continuous and satisfy the equality (19). Theorem 4 is proved.
 
The function of a set $\nu $ is defined on the class $Cyl _{2}$ only since the convergence of semigroups sequence can't obtain the limit behavior of sequence of the values of pseudomeasures $\mu _{n}$ on the arbitrary sets of the algebra ${\cal A}$. In addition if  the semigroup sequence $\{ {\bf U}_n\} $ converges in the weak operator topology uniformly on arbitrary segment only then the equality (17) should not be fulfilled.
 
In fact, if $A^{0,t,t+s}_{V,W,Y}\in
 Cyl_{3}\subset {\cal A}$ is the cylindrical set with three dimensional base then for any $n\in
 \bf N$ the Markovian pseudomeasures $\mu _n$ satisfy the equality (17). But according to the work \cite{STMF} (or \cite{SR11}) the following inequality $\lim\limits_{n\to
 \infty}({\bf U}_{n}(s)\chi _{Y},{\bf P}_{W}{\bf U}_{n}(t)\chi
 _{V})\neq ({\bf F}(s)\chi _{Y},{\bf P}_{W}{\bf F}(t)\chi _{V})$ holds. Moreover, the numerical sequence $\{ ({\bf U}_{n}(s)\chi
 _{Y},{\bf P}_{W}{\bf U}_{n}(t)\chi _{V}) \}$ can diverges fore some choice of the setsn $Y,V,W\in {\cal
 B}_{f}(R^{d})$.

\section { Random semigroups}. The following extension of notion of random variable is introduced. {\it Random variable is defined as the measurable map of the space with finite additive measure $(\Omega ,{\cal F},\mu )$ into the measurable space $(Y,{\cal A})$.} The random variable with the value {\it in topological space} $(Z,\tau )$ is defined as  the measurable map into the measurable space $(Z,{\cal A}_{\tau })$ where algebra of (Borel) subsets ${\cal A}_{\tau }$ is {\it the least algebra containing the topology $\tau$} (i.e. containing any open subset of the space $(Z,\tau )$).

Let $E$ be some topological space and $X$ is some Banach space which has the predual space $X_*$. The set $E$ can be presented by the set $G(X)$ of all generators of strongly continuous semigroups of transformations of Banach space $X$ which is endoved with the topology of strong (or weak) graph convergence.

According to our notations the symbol $W(E)$ notes the set of nonnegative normalized finite additive measures on the measurable space $(E,2^E)$. For any $\e _0\in E$ the symbol $W(E,\e _0)$ notes the set of measures  $\{ \mu \in W(E)$ such that 1) $\mu (A)=0$ if the cloasure $\bar A$ of the set $A\subset E$ in the topological space $E$ does not contain the point $\e _0$;  2) $\mu (\{ \e _0\})=0$.

Let $(E,2^E,\mu )$ be the measurable space endowing with the measure $\mu \in W(E)$. Let $\xi $ is random variable which is defined on the set $E$ and takes values in the topological space $Z_{s(w)}=C_{s(w)}(R_+,B(X))$ of strongly (weakly) continuous maps of semiaxe $R_+$ into the Banach space  $B(X)$ of bounded linear operators in the space $X$. 

Let us consider a family of functionals $\{ \f _{t,A,g},\ t\in R_+,A\in X,g\in X_* \}$ on the space $Z_w$ such that  $\f _{t,A,g}(z)=\sup\limits_{\t \in [0,t]}\langle z(\t )A,g\rangle ,\ t\in R_+,A\in X,g\in X_*$ for any $z\in Z_w$. Let $\tau _w$ be the toplogy on the linear space $Z_w$ which is generated  by the family of functionals $\f _{t,A,g},\ t\in R_+,A\in X,g\in X_*$.  

The space $Z_s$ can be endowed with the family of the functionals  $\{ \Phi _{t,A},\ t\in R_+,A\in X\}$ such that  $\Phi _{t,A}(z)=\sup\limits_{\t \in [0,t]}\sup\limits_{g\in X_*,\, \|g\|_{X_*}=1}\langle z(\t )A,g\rangle ,\ t\in R_+,A\in X$ for any $z\in Z_s$. This family of functionals generates the topology $\tau _s$ on the space $Z_s$.

Then topological vector space $Z$ ($Z_s$ or $Z_w$) endowing with the algebra of Borel subsets ${\cal A}_{\tau }$ is the measurable space and the map $\xi :\ E\to Z$ is measurable. Therefore $\xi $ is random variable.

{\it A random variable $\xi :\ E\, \to \, Z_s$ ($\xi :\ E\, \to \, Z_w$) is called random semigroup iff its values $\xi (\e ),\, \e \in E,$ are stron (weak) continuous semigroups}.

Then the linear space $Z$ ($Z_s$ or $Z_w$) endowing the structure of algebra of Borel subsets is the measurable space and the map $\xi :\ E\to Z$ is the random variable.

The mean value of random variable $\xi $ as the map of the space with the measure $(E,2^E,\mu )$ into the linear topological space $Z$ is Pettis integral 
$$
M\xi =\int\limits_E\xi _{\e } d\mu (\e  ),
$$
where $M\xi $ is the element of the space $Z$ such that  the equality 
$$
\langle M\xi (t)A,g\rangle =\int\limits_E\langle \xi _{\e } (t )A,g\rangle d\mu (\e)\eqno (20)
$$ 
holds for any $t\in R_+,A\in X,g\in X_*$. Here the integral in right hand side of equality (20) is Radon integral of bounded measurable complex-valued function on the set $E$ by the finite additive measure $\mu $.

{\bf Lemma 2.} For any $t\in R_+$ the family of equalities (20) with arbitrary $A\in X,g\in X_*$ uniquely defines the element $M\xi (t)\in B(X)$.

In fact the Radon integral $\int\limits_E\langle \xi _{\e } (t )A,g\rangle d\mu (\e)$  of bounded measurable complex-valued function on the set $E$ by the finite additive nonnegative normalized measure $\mu $ exists and it is the bounded belinear function on the space $X\times X_*$.

{\bf Theorem 5.} {\it If there is the dense subset $D\subset X$ in the space $X$ such that fo any $A\in D$ the family of maps $\xi _{\e }(t)A \in C(R_+,X),\, \e \in E$ is weak (or strong) uniformly continuous and the family of maps $\xi _{\e},\, \e \in E$, is uniformly bounded then  for any measure $\mu \in W(E)$ the mean values of random variable $\xi $ is continuous operator-valued function: $M\xi  \in C_w(R_+,B(X))$ (or  $M\xi  \in C_s(R_+,B(X))$).       }

Proof. 
The weak uniformly continuity means that for any $A\in D$, any $g\in X_*$ and any $\s >0$ there is the constant  $\delta >0$ such that $\sup\limits_{t\in R_+,\e \in E}| \langle \xi _{\e }(t+\Delta t)A-\xi _{\e }(t)A, g \rangle |\leq \s $ if $ |\Delta t|<\delta $. The strong analog of this condition is following: for any  $A\in D$ and any $\s >0$ there is the constant  $\delta >0$ such that $\sup\limits_{t\in R_+,\e \in E}| \| \xi _{\e }(t+\Delta t)A-\xi _{\e }(t)A \|_X\leq \s $ if $|\Delta t|<\delta $. 

The  uniformly boundedness of random variable $\xi $
means that there is the constant $C>0$ such  that $\sup\limits_{\e \in E,\, t\in R_+}\| \xi _{\e }(t)\|_{B(X)}\leq C$.

Since the random variable $\xi $ is uniformly bounded then for any   $t\geq 0$, any $A\in X$ and any $g\in X_*$  the function  $\langle \xi _{\e }(t)A, g \rangle ,\, \e \in E,$ is bounded. Therefore for any measure $\mu \in W(E)$ the integral (20) is correctly defined as the Radon integral (see \cite{SR11}). Moreover the integral (20) as the fuction of  argument $g$ is the bounded linear functional on the space  $X_*$. Hence for any  $A\in X$ Pettis integral $\int\limits_{G(X)}\xi _{\e }Ad\mu (\e )\in X$ is correctly defined. 
Then for arbitrary $t>0$ the mean value $M\xi (t)= \int\limits_{G(X)}\xi _{\e }d\mu (\e ) \in B(X)$ is correctly defined. 

According to the weak uiformly continuity condition for any $A\in D$, any $g\in X_*$ and any $\s >0$ there is the constant  $\delta >0$ such that for any $t\in R_+$ and any $\Delta t\in (0,\delta ):\, t+\Delta t\in R_+$ the following estimates take place $\sup\limits_{t\in R_+}|\langle (M\xi (t+\Delta t)-M\xi (t))A,g\rangle |=\sup\limits_{t\in R_+}| \langle \int\limits_{E}[\xi _{\e }(t+\Delta t)A-\xi _{\e }(t)A, g \rangle ]d\mu |\leq \int\limits_{E} \sup\limits_{t\in R_+,\e \in E}| \langle \xi _{\e }(t+\Delta t)A-\xi _{\e }(t)A, g \rangle |d\mu \leq \s $. 

If the strong uiformly continuity condition holds then for any $A\in D$ and any $\s >0$ there is the constant  $\delta >0$ such that for any $t\in R_+$ and any $\Delta t\in (0,\delta ):\, t+\Delta t\in R_+,$ the following estimates take place $\sup\limits_{t\in R_+} \| M\xi (t+\Delta t)A-M\xi (t)A\|_X=\sup\limits_{t\in R_+}\sup\limits_{\|g\|_{X_*}=1}|\langle (M\xi (t+\Delta t)-M\xi (t))A,g\rangle |\leq \int\limits_{E} \sup\limits_{t\in R_+,\e \in E}\sup\limits_{\|g\|_{X_*}=1}| \langle \xi _{\e }(t+\Delta t)A-\xi _{\e }(t)A, g \rangle |d\mu =\int\limits_{E} \sup\limits_{t\in R_+,\e \in E} \| \xi _{\e }(t+\Delta t)A-\xi _{\e }(t)A \| _Xd\mu\leq \s $.

For any $u\in X$ there is an element $A\in D $ such that $\|u-A\|_X\leq \s $. Then according to uniformly boundedness condition the easimate $\| M\xi (t)u-M\xi (t)A\|_X\leq C\s $ holds for any $t\geq 0$. Thus the continuity of  the mean value $M\xi $ of random variable $\xi $ takes place in corresponding topologies.

{\bf Remark.} The examples of families of operetor-valued functions from the space $C_s(R_+,B(X))$ which satisfies the condition of dense weak (strong) uniformly continuousness are considered in the papers \cite{IDA, SR11}.

{\bf Corollary 4.} Suppose that $S\subset E$ and $s_0\in E$ is the limit poin of the set $S$. Let $\xi $ be the map of the set $S$ into the topological vector space $C_w(R_+,B(X))$ such that for any $s\in S$ the value $\xi (s)={\bf U}_s\in C_w(R_+,B(X))$ is strongly continuous one-parameter semigroup.

I) If the map $\xi :\ S\to C_w(R_+,B(X))$ has the limit  $\xi _0$ as $s\to s_0$ then for any measure $\mu \in W(S,s_0)$ the mean value of random variable $\xi :\ (S,2^S,\mu )\to  C_w(R_+,B(X))$ coincides with the limit  $\xi _0$.

II) If the image $\Xi$ of the map $\xi :\ S\to 
C_w(R_+,B(X))$ is precompact then for any measure  $\mu \in W_0(S,s_0)$ the  mean value of random variable $\xi :\ (S,2^S,\mu )\to  C_w(R_+,B(X))$ is the limit of the map  $\xi $ in the point $s_0$ with respect to ultrafilter $\digamma _{\mu }=\mu ^{-1}(1)$. Conversely for any limit point $\xi _0\in C_w(R_+,B(X))$ of the image $\Xi$ there is the measure $\mu _0\in W_0(S,s_0)$ such than  $\xi _0$ is the mean value of the random variable $\xi :\ (S,2^S,\mu _0)\to  C_w(R_+,B(X))$.

The statement of the corollary 4 is the consequence of theorems 1 and 5. The examples of applications of corollary 4 to concreet maps is given in the papers \cite{SR11}. 

Thus under the assumptions of the theorem 5 or corrolary 4 the  mean value of the random variable $\xi $ is an element of the space $C_w(R_+,B(X))$ which will be called as the family of  averaging maps of the space $X$.

{\bf Remark 7}. The mean value of random semigroup can has no semigroup property. The trivial example of this phenomenon is the averaging of two unitary semigroup  $e^{-it},\, t\geq 0$ and $e^{it},\, t\geq 0$, acting in one-dimentional complex Hilbert space $\bf C$. If each of this semigroup has the probability ${1\over 2}$ then the mean value of this random semigroup is one-parameter family $F(t),\, t\geq 0,$ of maps of the space $\bf C$, any of each acting as the multiplier on the number $\cos t$.  
Since $\cos (t+s)\neq (\cos t)(\cos s)$ then the family of averaging maps $F(t),\, t\geq0,$ is not semigroup.
The paper \cite{Tartar} containes the example of the sequence of approximating semigroup which has the limit in the weak operator topology. The limit operator valued function for this sequence of semigroups has no semigroup property and has memory effect of dependence of the derivative on the values of the function in previos moments of time. Since limit of the sequence is the mean value of random semigroups on the measurable space with espetial measure then the mean value of random semigrops can possesses the memory effects.

\section
{The pseudomeasures associated with the Cauchy problem}. 
Let symbol ${\cal L}(C(R_+,R^d),{\cal A},\s _{cyl})$ note the topological vector space of pseudomeasures on the measurable space 
$(C(R_+,R^d),{\cal A})$ endowed with the topology $\s _{cyl}$ which is generated by the cylindrical functionals. (The definition of topology $\sigma _{cyl}$ is given by the equality (12)). Any map $\xi \ E\, \to \,  {\cal L}(C(R_+,R^d),{\cal A},\s _{cyl})$ is the measurable map of the space with the measure $(E,2^E,\mu )$ into the measurable space $({\cal L}(C(R_+,R^d),{\cal A},\s _{cyl}),{\cal A}_{\s _{cyl}})$. {\it Any random variable $\xi :\ E\, \to \, {\cal L}(C(R_+,R^d),{\cal A},\s _{cyl})$ is called random pseudomeasure}. 

Any continuous pseudomeasure $\nu $ on the algebra ${\cal A}$ (which variation can be infinite) generates the two-paramater family of operators ${\bf V}^{\mu }_{t_1,t_2},\, t_1,t_2\in R_+,$ in the Banach space $B(H)$ which is defined by the family of bounded bilinear forms $\beta _{\mu ,t_1,t_2}$ (пїЅпїЅ. (6), (10)).
Conversely any two-parameter family of bounded linear operators ${\bf V}_{t_1,t_2},\, t_1,t_2\in R_+,$ in Hilbert space  $H$ defines the Markovian bounded pseudomeasure $\nu _{{\bf V}}$ on the algebra ${\cal A}$ (which variation can be infinite) whih is given by the equality (3).

Therefore for any random variable $\xi _{\e }$ with the values in topological vector space $Z$ the random variable  $\nu _{\e }$ with the values in topological vector space ${\cal L}(C(R_+,R^d),{\cal A},\s _{cyl})$ corresponds in accordance with (3). 

Moreover the convergence of the sequence of operators 
${\bf \xi }_{\e },\, \e \in E,$ as $\e \to \e _0$ in the strong (weak) operator topology is connected with the converence of the sequence of pseudomeasures in the space   ${\cal L}(C(R_+,R^d),{\cal A},\s _{cyl})$ by the theorems 2, 3. 

\subsection
{The averaging procedure for the sequence of pseudomeasures.} The sequence of the values of the measures $\mu _{n}$ on some elements of the famly of the cylindrical subspace $A=A^{t_{1},...,t_{n}}_{B_{1},...,B_{m}},\, m\in {\bf N}$ can diverges. To investigate the limit bechavior of the sequence of pseudomeasures $\mu _{n}$ and to obtain the limit function of a set on the algebra  ${\cal A}$ the methods of the Banach limits and invariant means (see  \cite{AS}, \cite{Bog}
\cite{Srinivas}, \cite{S06}, \cite{STMF}, \cite{IDA}) have being used. For this aim the measurable space of regularizing parameters $({\bf N},2^{{\bf N}})$ endowed with the nonnegative normalized finitely additive measure $\varrho \in ba({\bf N},2^{{\bf N}})=l_{\infty }^{*}$, which is concentrated in the arbitrary neighborhhod of the point $\infty =\sup ({\bf
N}) $. The existence of a nontrivial class of mesures with  the above property and its studying had beingn published  in the papers \cite{IH, S06}.

The procedure of the averagind ofthe sequence of pseudomeasures $\mu _{n}$ with values in the space
${\cal L}(C(R_{+},R),{\cal A})$ by measure $\varrho $ is given by the Pettis integral. The averaging function of a set $\mu ^{\varrho
}$ on the algebra ${\cal A}$ is given by the relation $\mu ^{\varrho
}=\int\limits_{{\bf N}}\mu _{n}d\varrho (n)$: 
$$ 
\mu
^{\varrho
}(A^{t_{1},...,t_{m}}_{B_{1},...,B_{m}})=\int\limits_{{\bf N}}\mu
_{n}(A^{t_{1},...,t_{m}}_{B_{1},...,B_{m}})d\mu (n) \ \forall \
A^{t_{1},...,t_{m}}_{B_{1},...,B_{m}}\in {\cal A}. 
$$
The averaging function of a set $\mu ^{\varrho}$ is defined on the algebra ${\cal A}$ and it is additive function,
i.e. $\mu ^{\varrho}$ is pseudomeasure.

{\bf Definition 5.}
The random variable with the values in the measurable space $({\cal L}(C(R_+,R^d),{\cal A}),\s _{cyl})$ of pseudomeasures is the measurable  map of the measurable space $(E,2^E)$ endowed with the measure $\nu \in W(E)$ into the topological vector space $({\cal L}(C(R_+,R^d),{\cal A}),\, \s _{cyl})$.

{\bf Definition 6.}
The random pseudomeasure $m:\ (E,2^E,\nu )\to {\cal L}(C(R_+,R^d),{\cal A},\s _{cyl})$ is called weakly iniformly bounded if for any $A\in {\cal A}$ there is the constant $M>0$ and the set  $E(A)\subset E$ such that $|m_{\e }(A)|\leq M$ for all $\e \in E(A)$ and $\nu (E(A))=1$.

Suppose that  $\nu \in W(E)$ and for any $\e \in E$ the pseudomeasure $\mu _{\e }$ is defined on the algebra ${\cal A}$ of subsets of the space $\Omega =C(R_+,R^d)$. Let the function of a set $\nu *\mu $ be defined  by the equality $\nu * \mu (A_E\times A_{\Omega })=\int\limits_{A_E}\mu _{\e }(A_{\Omega })d\nu (\e )$ for arbitrary $A_E\in 2^E$ and $A_{\Omega }\in {\cal A}$.

{\bf Lemma 3.} 
{\it If  $\nu \in W(E)$ and random pseudomeasure $\mu $ on algebra ${\cal A}$ is weakly uniformly bounded,
then the function of a set $\nu *\mu $ is defined on the algebra $2^E\otimes {\cal A}$ and is finite additive.}

In fact for any $A_{\Omega }\in {\cal A}$ the numerical function $\mu _{\e }(A_{\Omega }),\, \e \in E,$ is defined on the set $E$ and is bounded on the set $E(A_{\Omega })$.    Since $\nu \in W(E)$ and $\nu (E(A_{\Omega }))=1$ then the function $\mu _{\e }(A_{\Omega }),\, \e \in E,$ 
is integrable with respect to measure $\nu $. 
Hence the function $\nu *\mu $ is defined on the following family sets  $\{ A_E\times A_{\Omega };\ A_{E}\in 2^E,\, A_{\Omega }\in {\cal A}\}$ and this function satisfies the properties of additivity with respect to both arguments. In fact, the additivity property of the function $\nu *\mu $ with respect to the first argument is the consequence of the additivity of Pettis the minimal algebra conteining the minimal algebra conteining the minimal algebra conteining integral. It's additivity with respect to the second argument is the consequence of linearity of Pettis integral and additivity of pseudomeasures  $\mu _{\e }$. 
Hence the function $\nu *\mu $ has the unique continuation by the standatr sceme onto the finite additive function of a set on the minimal algebra conteining the family sets  $\{ A_E\times A_{\Omega };\ A_{E}\in 2^E,\, A_{\Omega }\in {\cal A}\}$,  i.e. on the algebra $2^E\times {\cal A}$.

{\bf Definition 7.} The mean value of random pseudomeasure $\mu $ on the space with the measure $(E,2^E,\nu )$ 
is defined as the pseudomeasure $\mu ^{\nu}$ on the algebra ${\cal A}$ such that 
$\mu ^{\nu }( A)=\int\limits_{E}\mu _{\e }(A)d\nu (\e )\ \forall \ A\in {\cal A} $.

Since the pseudomeasure $\mu ^{\nu}$ is equal to the restriction of pseudomesure $\nu *\mu $ onto the subalgebra $E\otimes {\cal A}$ then the pseudomeasure $\nu *\mu $ is Young pseudomeasure with respect to pseudomeasure $\mu ^{\nu }$ in the sense of definition 3.5.7 of  \cite{NA}.

{\bf Lemma 4.} {\it Suppose that $\nu \in W(E)$.
If the random pseudomeasure ${\mu }$ on the space $(E,2^E,\nu )$ is weakly uniformly bounded
then  it's mean value  is the element of the space ${\cal L}(C(R_+,R^d),{\cal A})$}.

In fact the pseudomeasure $\nu ^{\mu } $ is defined on the algebra ${\cal A}$ and is additive function on this algebra.
It should be note that the pseudomeasure $\nu ^{\mu }$
is equal to the restriction of pseudomeasure $\nu *\mu $ on the subalgebra $E\otimes {\cal A}$.

{\bf Remark 8}.
{ If for any $\e \in E$ the pseudomeasure $\mu _{\e }$ is continuous (see definition  1), then the family of maps  ${\bf V}_{\mu _{\e }}$  is the family of bounded linear operators in the space $X$.  
Moreover if for any  $\e \in E$ the pseudomeasure $\mu _{\e }$ is Markovian and stationar  then the family of maps ${\bf V}_{\mu _{\e }}$ is oneparameter semigroup of bounded linear operators in the space $X$.}
{However even the measure $\mu _{\e}$ is continuous, stationary  and Markovian for any $\e \in E$ but the averaging measure  $\mu ^{\nu }$ can has no Markovian property and has no continuity property.}

\section
{Topologies on the space of pseudomeasures geterated by the nonlinear cylindric functional.
}
The topology in the space of Markovian continuous pseudomeasures  (see \cite{SS2}) which is corresponded by the convergence of semigroup in the space $H$ %
is generated by the family of the cylindrical functional  $P_{A}(\mu ),\, \mu \in
{\cal L}(C(R_{+},R^{d}),{\cal A})$ which is paremetrized by the family of sets $\{ A\in  {\cal A}\}$.
In particular if the functional $P_{A}$ is given by some $ A\in Cyl_{2}$ then the linear functionals $P_A$ is given by (13).

Let us introduce the nonlinear functionals on the space of pseudomeasures which is defined by using of quantum states.
The symbol $\Sigma (H)$ notes the set of quantum states (see \cite{AS} \cite{PV}, \cite{Sizv}) which is the intersection of the unite sphere with the positive cone of Banach space 
$(B(H))^*$ (here $(B(H))^*$ is conjugate to the space $B(H)$ of bounded linear operators).

If $\mu $ is continuous pseudomeasure $\mu \in {\cal L}_{cont}(C(R_+,R^d),{\cal A})$ then for any nonnegative numbers $t_{1},t_{2}\in R_+$ the bounded sesquilinear form $\beta _{\mu }^{t_1,t_2}$ defines two-parametric family of bounded linear operators ${\bf U}_{\mu }^{t_0,t_1},\, t_0,t_1\in R_+,$ which is given by tne pseudomeasure $\mu $ in accordance with the equalities  (6). 
For any bounded linear self-adjoint operator ${\bf A}$ with the descrete spectrum $\sigma ({\bf A})=\{ a_{k}\}$ which has the orthonormal basis $\{ \psi _{k}\}$ of eigen vectors,  for any $u\in H$ and any $t_{1},t_{2}\in R_+$ the following number is defined
$$
F^{t_{0},t_{1}}_{\rho _u,{\bf A}}(\mu )=
\sum\limits_{k=1}^{\infty }a_{k}| \int\limits_{C(R_{+},R)}u(\xi
(t_{0}))\bar \psi _{k}(\xi (t_{1}))d\mu (\xi ) |^{2}=\sum\limits_{k=1}^{\infty }a_{k}| (u,{\bf U}_{\mu }^{t_0,t_1}\psi _k) |^2
$$
for arbitrary continuous measure $\mu \in {\cal L}_{cont}(C(R_+,R^d),{\cal A})$. 

Since the set of bounded linear operators with the descrete spectrum is dense in the space of bounded linear operators  then for any  $t_0,t_1\geq 0,\, \mu \in {\cal L}_{cont}(C(R_{+},R^{d}),{\cal A})$ and $u\in H,\, \|u\|=1$ the functional  $F_{\rho _u,{\bf \cdot}}^{t_0,t_1}(\mu )=\langle \rho _u,{\bf U}_{\mu }^{t_0,t_1}{\bf \cdot}({\bf U}_{\mu }^{t_0,t_1})^*\rangle $ can be defined according to continuity on the hhole space $B(H)$. Since any element $\rho \in \Sigma (H)$ can be decomposed into the Pettis integral  on the set of pure vector states (see \cite{AS, Sizv}) then for any fixed values $t\geq 0$, ${\bf A}\in B(H)$ and $\mu \in {\cal L}_c(C(R_{+},R^{d}),{\cal A})$ the functional $F_{\cdot ,{\bf A}}^{t_0,t_1}(\mu )$ can be continued according to linearity on the hole set $\Sigma (H)$ by the rule: if the state $\rho \in \Sigma (H)$ cab ne docomposed by Pettis integral $\rho =\int\limits_{\|e\|=1}\rho _{e}d\lambda (e)$ with respect to some nonnegative normalised finite additive measure $\lambda $ on the unite sphere $(S_1(H),2^{S_1(H)})$ of Hilbert space $H$ then the value of functional $F_{\rho ,{\bf A}}^{t_0,t_1}$ on the pseudomeasure $\mu \in {\cal L}_{cont}(C(R_{+},R^{d}),{\cal A})$ is equal $F_{\rho ,{\bf A}}^{t_0,t_1}(\mu )=\int\limits_{\|e\|=1}F_{\rho _e,{\bf A}}^{t_0,t_1}(\mu )d\lambda (e)$. 

Thus if $\mu  \in {\cal L}_{cont}(C(R_{+},R^{d}),{\cal A})$ is continuous pseudomeasure generating the two-parameter family ${\bf U}_{\mu }^{t_1,t_2},\, t_1,t_2\in R_+,$ of bounded linear operators then for any  ${\bf A}\in B(H)$, $\rho \in \Sigma (H)$; $t_1,t_2\in R_+$ the following equality holds
$$
F_{\rho ,{\bf A}}^{t_1,t_2}(\mu )=(\rho ,{\bf U}_{\mu }(t_2-t_1){\bf A}{\bf U}^{-1}_{\mu }(t_2-t_1)),\quad \forall \ t_1,t_2:\  t_2\geq t_1\geq 0,\, \rho \in \Sigma (H),\, {\bf A}\in B(H).\eqno (21)
$$

For any  $\rho \in \Sigma (H)$,  $t_{1},t_{2}\in R_+$ and ${\bf A}\in B(H)$ the functional (21)
$F^{t_{0},t_{1}}_{\rho ,{\bf A}}$ is the quadratic functional on the space ${\cal L}(C(R_{+},R^{d}),{\cal A})$, which is defined on some subset of the space ${\cal L}(C(R_{+},R^{d}),{\cal A})$. 

The symbol $\cal F$ notes the family of functionals $\{ F^{t_1,t_2}_{\rho , {\bf A}},\ t_1,t_2\in R_+,\ \rho \in \Sigma (H),\, {\bf A}\in B(H)\}$. Fny functional of this family is defined on some subset of linear space ${\cal L}(C(R_{+},R^{d}),{\cal A})$.

The symbol $\cal D$ notes the set ${\cal D}\subset {\cal L}(C(R_{+},R^{d}),{\cal A})$ of pseudumeasures of the space ${\cal L}(C(R_{+},R^{d}),{\cal A})$ which benongs to the definition domain of all functionals of the family $\cal F$.
As it is shown above the family of the functionals $\{ F^{t_1,t_2}_{\rho ,{\bf A}};\ t_1,t_2\in R_+,\ \rho \in \Sigma (H),\, {\bf A}\in B(H)\}$ is defined on the space ${\cal L}_{cont}(C(R_+,R^d),{\cal A})$, i.e. $
{\cal L}_{cont}(C(R_+,R^d),{\cal A})\subset {\cal D}$.
On the other side if $\mu \in {\cal L}(C(R_{+},R^{d}),{\cal A})$ is some pseudomeasure  such that any functional (21) is defined on this pseudomeasure
then the bilinear form $\beta ^{t_0,t_1}_{\mu }$ must be bounded on the space $H_S$ for arbitrary $t_0,t_1\geq 0$. Therefore the set $\cal D$ is the set of pseudomesures satisfying the conditions (5a) for $m=1$ and any $t_0,t_1\geq 0$.

{\bf Lemma  5.} If $\mu ,\nu \in {\cal D}$ then $\mu +\nu \in {\cal D}$ and $\a \mu \in {\cal D}$ for any $\a \in {\bf C}$.

Since  $\int\limits_{C(R_+,R)}u(\xi (t_0))\bar \psi _k(\xi (t_1))d(\mu +\nu )(\xi )= \int\limits_{C(R_+,R)}u(\xi (t_0))\bar \psi _k(\xi (t_1))d(\mu )(\xi )+\int\limits_{C(R_+,R)}u(\xi (t_0))\bar \psi _k(\xi (t_1))d(\nu )(\xi )$ then the following inequality holds $|\int\limits_{C(R_+,R)}u(\xi (t_0))\bar \psi _k(\xi (t_1))d(\mu +\nu )(\xi )|^2\leq 2(|\int\limits_{C(R_+,R)}u(\xi (t_0))\bar \psi _k(\xi (t_1))d(\mu )(\xi )|^2+|\int\limits_{C(R_+,R)}u(\xi (t_0))\bar \psi _k(\xi (t_1))d(\nu )(\xi )|^2)$. Thus for any values $t_1,t_2,{\bf A},\rho $ and any pseudomeasures $\mu ,\nu \in {\cal D}$ the following inequality holds:
$$
F^{t_1,t_2}_{\rho ,{\bf A}}(\mu +\nu )\leq 2(F^{t_1,t_2}_{\rho ,{\bf A}}(\mu )+F^{t_1,t_2}_{\rho ,{\bf A}}(\nu )).\eqno (22)
$$  
Therefore the values of the functionals $F^{t_1,t_2}_{\rho ,{\bf A}}(\mu +\nu )$ are defined for any  $t_1,t_2\in R_+,\ \rho \in \Sigma (H),\, {\bf A}\in B(H)$ because the values of functionals $F^{t_1,t_2}_{\rho ,{\bf A}}(\mu )$ and $F^{t_1,t_2}_{\rho ,{\bf A}}(\nu )$ are defined for any $t_1,t_2\in R_+,\ \rho \in \Sigma (H),\, {\bf A}\in B(H)$.

The second statement of the lemma is the consequence of the equality $F^{t_1,t_2}_{\rho ,{\bf A}}(\a \mu )=\| \a \|^2F^{t_1,t_2}_{\rho ,{\bf A}}(\mu )$ which holds according to  (22) for any values of $t_1,t_2,{\bf A},\rho $.

{\bf Corollary 5}. The set ${\cal D}$ is the linear subspace in the space ${\cal L}(C(R_+,R^d),{\cal A})$ such that ${\cal L}_{cont}(C(R_{+},R^{d}),{\cal A})\subset {\cal D}\subset {\cal L}(C(R_{+},R^{d}),{\cal A})$.

Let the topology $\tau _{\cal F}$ on the linear space $\cal D$ be defined by the family of functionals  ${\cal F}$ in the following sence: the topology $\tau _{\cal F}$ is the least toplogy containing the sets $\{ f^{-1}(B),\ f\in B,\, B\in {\cal B}({\bf C})\}$ where ${\cal B}({\bf C})$ is Borel algebra (but not only system of open sets) of subsets of comlex plane $\bf C$. If the set $A_{f,B}\in \tau _{\cal F}$ for some $f\in {\cal F},\, B\in {\cal B}({\bf C})$ is defined by the condition $A_{f,B}=\{ \mu \in {\cal D}:\ f(\mu )\in B\}$ then the set ${\cal D}\backslash A_{f,B}=\{ \mu \in {\cal D}:\ f(\mu )\notin B\}=A_{f,R^d\backslash B}$ belongs to the topology $\tau _{\cal F}$. Thus the topology $\tau _{\cal F}$ is closed with respect to complement operation (any open set of the toplogy $\tau _{\cal F}$ is closed set).

Let symbol $\Sigma  _{\cal F}$ note the algebra of subsets of the space $\cal D$ which is generated by the family of functionals $\cal F$ i.e.  $\Sigma  _{\cal F}$ is the least algebra containing the sets $\{ f^{-1}(B),\ f\in {\cal F},\, B\in {\cal B}({\bf C}) \}$. Then the following condition $\Sigma _{\cal F}\subset \tau _{\cal F}$ holds because the topology $\tau _{\cal F}$ is closed with respect to complement operation. 

{\bf Remark 9}. The  topolotgical subspace $({\cal L}_{cm}(C(R_{+},R^{d}),{\cal A}), \tau _{\cal F})$ of continuous Markovian pseudomeasures in the topological space $({\cal D},\tau _{\cal F})$ is Hausdorff space. In fact, due to Markovian property (5.6) there is the one-to-one correspondence between the set ${\cal L}_{cm}(C(R_{+},R^{d}),{\cal A})$ and the set of two-parametric families of bounded linear operators subjecting to dynamical property ${\bf U}^{t_2,t_3}{\bf U}^{t_1,t_2}={\bf U}^{t_1,t_3},\ 0\leq t_1\leq t_2\leq t_3$.And if $(\rho ,{\bf U}_{\mu _1}^{t_1,t_2}{\bf A}({\bf U}^{t_1,t_2}_{\mu _1})^*)=(\rho ,{\bf U}_{\mu _2}^{t_1,t_2}{\bf A}({\bf U}^{t_1,t_2}_{\mu _2})^*)$ for all  $t_1,t_2:\  t_2\geq t_1\geq 0,\, \rho \in \Sigma (H),\, {\bf A}\in B(H)$, then $\mu _1=\mu _2$ because any Markovian pseudomeasure can be uniquely reconstructed by its restriction on the class ${\rm Cyl}_2$ of cylindrical subsets with two-dimentional base. But the hole space  $({\cal D},\tau _{\cal F})$ is not Hausdorff space since there is some two continuous (but not Markovian) pseudomeasures $\mu _1,\mu _2$ such that it have the common restrictions on the classпїЅ  ${\rm Cyl}_2$ and distinguishing restrictions on the class ${\rm Cyl}_3$.

For any pseudomeasure $\mu _0\in {\cal D}$ the following class of pseudomeasures $K_{\mu _0}=\{ \mu \in {\cal D}:\ f(\mu )=f(\mu _0)\ \forall \ f\in {\cal F}\}$ is defined. The class of pseudomeasures $K_{\theta }=\{ \mu \in {\cal D}:\ f(\mu )=0\ \forall \ f\in {\cal F}\}$ is the linear subspace in the linear ${\cal D}$ according to the lemma 5 (see 22)). Hence the factor-space ${\cal D}/ K_{\theta }$ is Hausdorff topological vector subspace in the space $\cal D$. Since any functional of the family $\cal F$ has the constant values on the sets $K_{\mu _0}, \mu _0\in {\cal D}$ then any own (proper) subset of a set $K_{\mu _0}, \mu _0\in {\cal D}$ is no element of the topology $\tau _{\cal F}$.
Hence any measurable numerical function on the measirable space $({\cal D},\Sigma _{\cal F})$ has the constant value on the sets  $K_{\mu _0}, \mu _0\in {\cal D}$. I.e. if ${\cal D}_{\theta }={\cal D}\backslash K_{\theta }$ then any measurable function on the space $({\cal D},\Sigma _{\cal F})$ takes constant values on the sets $K_{\mu _0}, \mu _0\in {\cal D}$. Therefore any  measurable function on the space $({\cal D},\Sigma _{\cal F})$ is the function on the factor-space $({\cal D}_{\theta },\Sigma _{\cal F})$.

Let symbol $(B({\cal D},\Sigma _{\cal F}),\ \tau _{D})$ note the topological vector space of measurable functions on the  topological vector space $({\cal D},\tau _{\cal F})$. The topology $\tau _{D}$ in the space $(B({\cal D},\Sigma _{\cal F}),\ \tau _D)$ is generated by the family of functionals $\f _{\mu },\, \mu \in {\cal D}$, any of each acting by the formular $\f _{\mu }(f)=f(\mu ),\ f \in B({\cal D},\Sigma _{\cal F})$. 
It should be noted that the functions of family $\cal F$ (but not only it) belong to the  topological vector space $(B({\cal D},\Sigma _{\cal F}),\ \tau _{D})$.

{\bf Lemma 6.} {\it If the topology $\tau _{B}$ on the linear space $\cal D$ is generated by the family of functionals $B({\cal D},\Sigma _{\cal F})$ then the topology $\tau _B$ (is equivalint to) coinsides with the topology $\tau _{\cal F}$.}

It is obvious that  $\tau _{\cal F}\subset \tau _B$.
If we asume  that the topology $\tau _B$ is more wide than the topology $\tau _{\cal F}$, then (according to definition of the space $B({\cal D},\Sigma _{\cal F})$) there is the measurable (with respect to the algebra   $\Sigma _{\cal F}$) functional $g\in B({\cal D},\Sigma _{\cal F})$ and there is the Borel subset $B\in {\cal B}({\bf C})$ such that  the set $g^{-1}(B)$ is not belong to the topology $\tau _{\cal F}$. But it is the contradiction to the measurability of the functional $g$ with respect to the algebra   $\Sigma _{\cal F}$ because $g^{-1}(B)\in \Sigma _{\cal F}$ for any  $B\in {\cal B}({\bf C})$ and   $\Sigma _{\cal F}\subset \tau _{\cal F}$. 
The proposition is proved.

Let us introduce the functional ${\cal V}$ on the linear space ${\cal D}$ which values is given by the formular ${\cal V}(\mu )=\sup\limits_{\|{\bf A}\|=1,\, t_1,t_2 \in R^+,\, \rho \in \Sigma (H)}|F^{t_1,t_2}_{\rho ,{\bf A}}(\mu )|$ for any $\mu \in \cal D$. The functional $\cal V$ is defined on the linear space $\cal D$ and it is the seminorm on the space   ${\cal D}$ according to the inequality (22). 
Then the functional ${\cal V}$ is the norm on the factor-space ${\cal D}_{\theta }={\cal D}\backslash K_{\theta}$.

Let symbol $b({\cal D},\Sigma _{\cal F})$  note the Banach space of bounded measurable functions on the space  $\cal D$ endowing with the norm $\|F\|=\sup\limits_{\mu \in {\cal D}:\ {\cal V}(\mu )\leq 1}|F(\mu )|$. Then ${\cal F}\subset b({\cal D},\Sigma _{\cal F})$ and the following statement holds.

{\bf Lemma 7.} {\it If the topology  $\tau _{b}$ on the linear space $\cal D$  is generated by the family of functionals $b({\cal D},\Sigma _{\cal F})$ then the topology $\tau _{b}$ coinsides with the topology  $\tau _{B}$.}

The topology $\tau _{b}$ is no wider than the topology $\tau _B$ because $b({\cal D},\Sigma _{\cal F})\subset B({\cal D},\Sigma _{\cal F})$.
On the other side for any function $f\in B({\cal D},\Sigma _{\cal F})$ and any interval $\Delta \in R$  there is the function $\f \in b({\cal D},\Sigma _{\cal F})$ (which can have only two different values on the set $f^{-1}(\Delta ) $ and its complement) and the interval $\Delta '\in R$ such that $f^{-1}(\Delta )=\f ^{-1}(\Delta ')$. Hence the topology  $\tau _{B}$ is no wider (stronger) than the topology $\tau _{b}$.

{\bf Remark 10}. If $\Phi \in B({\cal D},\Sigma _{\cal F})$ (i.e. $\Phi ^{-1}(B)\in \Sigma _{\cal F}$ for any $B\in {\cal B}({\bf C})$), then $\Phi (\mu )=\Phi (\mu _0)\ \forall \ \mu \in K_{\mu _0}$.

Let symbol $ba ({\cal D},\Sigma  _{\cal F})$ note the Banach space of measures with finite variation on the measurable space $({\cal D},\Sigma  _{\cal F})$, i.e. the conjugate space for Banach space  $b({\cal D},\Sigma _{\cal F})$ (see \cite{DSh}).
The measure $m\in ba ({\cal D},\Sigma _{\cal F})$ is called as the limit point of the sequence of the measures $\{ m_n\}$ in the space $ba ({\cal D},\Sigma _{\cal F})$ if for any $\epsilon >0$ and any $F\in b({\cal D},\Sigma _{\cal F})$ the inclusion $F(m_n)\in O_{\epsilon }(F(m))$ holds for any $m$ from some infinite subset ${\bf N}_{\epsilon ,F}$ of the set  $\bf N$.

Let $E=(0,1)$ be the set of regularization parameters  and $\xi $ is the map $\xi:\ E\to {\cal D}$ such that the value $\xi (\e )$ on any point $\e \in E$ is the pseudomeasure $\mu _{\e }$ on the algebra ${\cal A}$ which is generated by the unitary semigroup ${\bf U}_{\e }$ in accordance with the equality (5). Then for any $\e \in E$ the value $\mu _{\e }$ of the map $\xi $ satisfies the condition $\mu _{\e }\in {\cal L}_{cm}(C(R_+,R^d),{\cal A})$ and hence  $\mu _{\e }\in {\cal D}$.

Let the measure $\delta (\mu -\mu _0)\in ba({\cal D},\Sigma _{\cal F})$ be defined by the condition: for any pseudomeasure $\mu _0\in {\cal D}$  the equality $\langle \delta (\mu -\mu _0),f\rangle =f(\mu _0)$ holds for any $f\in b({\cal D},\Sigma _{\cal F})$. Since the function $f\in b({\cal D},\Sigma _{\cal F})$ is measurable then it has the constant values on the sets $K_{\mu _0},\, \mu_0\in \cal D$. Therefore $\delta (\mu -\mu _0)=\delta (\mu -\mu _1)$ if $\mu _1\in K_{\mu _0}$.

{\bf Theorem 6.} {\it The set $M\subset ba ({\cal D},\Sigma _{\cal F})$ of values of the sequence of mesures $\delta (\mu -\mu _{\e }),\, \e \in E,$ is compact in the topology $\tau _{b}$.

If $\hat \nu  \in ba({\cal D},\Sigma _{\cal F})$ is the limit point of the sequence of measures  $\delta (\mu -\mu _{\e }),\, \e \in E,$ in the space $ba ({\cal D},\tau _{b})$
then   the equaliny  $F(\mu _{\e })\to \langle F(\mu )\rangle _{\nu }=\int\limits_{{\cal L}}F(\mu )d\nu (\mu )$ holds for any functional  $F\in b({\cal D},\Sigma _{\cal F} )$.}

The statement of this theorem is the realization of the theorem 1 statement for the spaces $Z=(b({\cal D},\Sigma {\cal F} ))^*=ba({\cal D},\Sigma {\cal F} )$ and $S=b({\cal D},\Sigma {\cal F} )$ and for the map $G:\ E\to Z$ acting by the equality $G(\e )=\delta (\mu -\mu _{\e }),\, \e \in E$.

In fact for any two-valued measure $\nu \in W_0(E)$ the functional $g(F)=\int\limits_{E}F(\mu _{\e })d\nu (\e )$ 
is linear and continuous functional on the space $b({\cal D},\Sigma {\cal F}  )$. Then the sequence of the measures $\{ \delta (\mu -\mu _{\e }),\, \e \in E\}$ converges by the ultrafilter $\digamma _{\nu }=\nu ^{-1}(1)$ to the measure $\nu $ in the following sence: for any functional $F\in b({\cal D},\Sigma {\cal F} )$ the equality $\lim\limits_{\e \to 0,\ \digamma _{\nu }}F(\mu _{\e }) = \langle F(\mu )\rangle _{\nu }=\int\limits_{E}F(\mu _{\e })d\nu (\e )$ holds.

Conversely 
if for some ultrafilter $\digamma $ of the set $E$ with the limit point $0$ the equality $\lim\limits_{\e \to 0,\, \e \in \digamma }F(\mu _{\e })=g(F)$ holds for any functional $F\in b({\cal D},\Sigma {\cal F} )$ then there is the mesure $\nu \in ba({\cal D},\Sigma _{\cal F})$ on the space ${\cal D}$ such that for any functional $F\in b({\cal D},\Sigma _{\cal F}  )$ the equality $g(F)\to \langle F(\mu )\rangle _{\nu }=\int\limits_{{\cal D}}F(\mu )d\nu (\mu )$ holds (the measure $\nu $ can be the image  under the action of the map $\xi :\ E\to {\cal D};\ \xi (\e )=\mu _{\e }$ of the measure on the set $E$ which is generated by the ultrafilter $\digamma $). Theorem 6 is proved. %

Thus the desctiption of the limit behavior of the class of continuous (not only linear) functionals on the space of pseudomeasures ${\cal D}$ 
is given by the measure $\nu $ on the space ${\cal D}$. In particular the description of the limit behavior  of continuous linear functionals of the space of pseudomeasures ${\cal D}$ (which is generated by the sequence of unitary groups in  Hilbert space of quantum system $H$) is given by the pseudomeasure $\mu _{\nu }\in {\cal D}$ which is barycenter of measure  $\nu $.

The measure $\nu $  on themeasurble  space of pseudomesures $({\cal L}_{cm}(C(R_+,\Omega ),\tau _{\cal F})$  is the analog of the family of Young mesures (see \cite{Pan, Sychev}). In fact let $\Omega \in R^d$ be some bounded domain with the smooth boundary. Let the strongly continuous semigroup of unitary operators in Hilbert space  $H=L_2(\Omega)$ be defined for any value of parameter $\e \in E=(0,1)$. Suppose that $\nu \in W_0(E)$. According to the equality (5) any semigroup ${\bf U}_{\e }$ of maps of the space $H$ defines the unique pseudomeasure $\mu _{\e}\in {\cal L}_{cm}(C(R_+,\Omega ))$. Suppose that there is the vector $u\in H$ such that for any $\e \in E$ and any $t\geq 0$ the function $u_{\e }(t)={\bf U}_{\mu _{\e}}(t)u$ is uniformly continuous on the domain $\Omega$ (the examples of such situations are described in the papers \cite{SR11, S06}).

Let us consider the class ${\cal F}_x$ of functionals $\{ \Phi _{t,x,\phi},\, (t,x)\in [0,T]\times \Omega,\, \phi \in C({\bf C})\}$ on the space $\cal D$ which are given by the equalities $\Phi _{t,x,\phi , u}=\phi (u_{\mu }(t,x))$ where $u_{\mu }(t,\cdot )=({\bf U}_{\mu }(t)u)(\cdot )$.
Then the measure  $\nu $ on the sequence of  pseudomeasures $\{ \mu _{\e }\},\, \e \in E,\, \e \to 0,$ uniquely defines the Young measure i.e. the family of measures $\{ \nu _{t,x},\, (t,x)\in [0,T]\times \Omega \}$ on the complex plane $\bf C$ such that for any $f\in C_0({\bf C})$ and any  $(t,x)\in [0,T]\times \Omega $ the equality $\lim\limits_{\e \to 0,\e \in \digamma _{\mu }}f(u_{\e }(t,x))= \int\limits_{E}f(u_{\e}(t,x))d\mu (\e) = \int\limits_{{\bf C}}f(\xi )d\nu _{t,x}(\xi )$ holds (see \cite{Sychev, S06}). Thus the measure on the space of pseudomeasures ${\cal L}_{c}(C(R_+,R^d),{\cal A})$ induces the Young measure corresponding to the sequence of approximations of solution of initial-boundary value problem.

\section*{Acknowledgment}

\quad The author thanks professors O.G.Smolyanov and N.N. Shamarov for their attention to this work and helpful advices.

This work is supperted by the grant RSF N 14-11-00687. 

\end{document}